\documentclass[traditabstract]{aa}
\usepackage{booktabs}
\usepackage{graphicx}
\usepackage[utf8]{inputenc}
\usepackage{natbib}
\usepackage{multirow}
\usepackage{rotating}
\usepackage{txfonts}
\bibpunct{(}{)}{;}{a}{}{,} % to follow the A&A style

\newcommand{\iraf}{\texttt{IRAF}}
\newcommand{\hst}{\textit{HST}}
\newcommand{\spitzer}{\textit{Spitzer}}
\newcommand{\swift}{\textit{Swift}}

\newcommand{\gcn}{GCN Circ.}

\title{Galaxy counterparts of intervening high-$z$ sub-DLAs/DLAs and \ion{Mg}{ii} absorbers 
towards gamma-ray bursts
\thanks{
Based in part on observations collected at the European Organisation for Astronomical Research
in the Southern Hemisphere, Chile, as part of the programs 075.A-0603, 075.A-0385, 077.A-0312,
084.A-0303, 177.A-0591 and 275.D-5022. Based in part on observations made with the NASA/ESA
\textit{Hubble Space Telescope}, obtained at the Space Telescope Science Institute, which is
operated by the Association of Universities for Research in Astronomy, Inc., under NASA contract
NAS 5-26555. Based in part on observations made with the \textit{Spitzer Space Telescope},
which is operated by the Jet Propulsion Laboratory, California Institute of Technology, under
a contract with NASA. Based in part on observations obtained at the Gemini Observatory, which
is operated by the Association of Universities for Research in Astronomy, Inc., under a
co-operative agreement with the NSF on behalf of the Gemini partnership.}}

\titlerunning{Galaxy counterparts of high-$z$ DLAs}

\author{
S.~Schulze\inst{1},
J.~P.~U.~Fynbo\inst{2},
B.~Milvang-Jensen\inst{2},
A.~Rossi\inst{3},
P.~Jakobsson\inst{1},
C.~Ledoux\inst{4},
A.~De~Cia\inst{1},
T.~Kr\"uhler\inst{2},
A.~Mehner\inst{4},
G.~Bj\"ornsson\inst{1},
H.-W.~Chen\inst{5},
P.~M.~Vreeswijk\inst{1},
D.~A.~Perley\inst{6},
J.~Hjorth\inst{2},
A.~J.~Levan\inst{7},
N.~R.~Tanvir\inst{8},
S.~Ellison\inst{9},
P.~M\o{}ller\inst{10},
G.~Worseck\inst{11},
R.~Chapman\inst{1,12},
A.~Dall'Aglio\inst{13},
G. Letawe\inst{14}
}
\offprints{S. Schulze, sts30@hi.is}

\institute{
  Centre for Astrophysics and Cosmology, Science Institute, University of Iceland, Dunhagi 3, 107 Reykjav\'ik, Iceland % Annalisa, Bob, Gulli, Palli, Paul, Steve
\and
  Dark Cosmology Centre, Juliane Maries Vej 30, 2100 K{\o}benhavn {\O}, Denmark % Bo, Daniele, Jens, Johan, Thomas
\and
  Th\"uringer Landessternwarte Tautenburg, Sternwarte 5, D-07778 Tautenburg, Germany % Andrea
\and
  European Southern Observatory, Alonso de Córdova 3107, Vitacura, Casilla 19001, Santiago 19, Chile % Cedric
\and
  Department of Astronomy, University of Chicago, 5640 S. Ellis Ave., Chicago, IL 60637, USA % Hsiao-Wen
\and
  Department of Astronomy, California Institute of Technology, MC 249-17, 1200 East California Blvd., Pasadena, CA 91125, USA % Dan
\and
  Department of Physics, University of Warwick, Coventry, CV4 7AL, United Kingdom % Andrew
\and
  Department of Physics and Astronomy, University of Leicester, Leicester LE1 7RH, United Kingdom % Nial
\and
  Department of Physics and Astronomy, University of Victoria, Victoria, British Columbia, V8P 1A1, Canada % Sarah
\and
  European Southern Observatory, Karl-Schwarzschildstrasse 2, 85748, Garching, Germany % Palle
\and
  UOC/Lick Observatory, University of California, 1156 High Street, Santa Cruz, CA 95064, USA % Gabor
\and
  Centre for Astrophysics Research, University of Hertfordshire, Hatfield, Herts AL10 9AB, UK % Bob
\and
  Astrophysikalisches Institut Potsdam, An der Sternwarte 16, 14482 Potsdam, Germany % Aldo
\and
  Institut d’Astrophysique et Géophysique, Université de Liège, Allée du 6 Août, 17 Bâtiment B5C, 4000 Liège, Belgium % Géraldine
}

\date{Received December 14, 2011; accepted XXXX}
 
\authorrunning{Schulze et al.}


\begin{document}

% \setpagewiselinenumbers
% \modulolinenumbers[5]
% \linenumbers

\abstract
{
We present the first search for galaxy counterparts of intervening high-$z$
($2< z < 3.6$) sub-DLAs and DLAs towards GRBs. Our final sample comprises of five
intervening sub-DLAs and DLAs in four GRB fields.
To identify candidate galaxy counterparts of the absorbers we use deep optical
and near-infrared imaging, and low-, mid- and high-resolution spectroscopy
acquired with 6 to 10-m class telescopes, the \textit{Hubble} and the \textit{Spitzer}
\textit{space~telescopes}. Furthermore,
we use the
spectroscopic information and spectral-energy-distribution fitting techniques
to study them in detail.
Our main result is the detection and spectroscopic confirmation of the galaxy 
counterpart of the intervening DLA at $z=3.096$ in the
field of GRB\,070721B ($z_{\rm GRB}=3.6298$) as proposed by other
authors. 
We also identify good candidates for the galaxy counterparts of the two strong \ion{Mg}{ii}
absorbers at $z=0.6915$ and 1.4288 towards GRB\,050820A ($z_{\rm GRB}=2.615$).
The properties of the detected DLA galaxy are typical for Lyman-break galaxies 
(LBGs) at similar redshifts; a young, highly starforming galaxy that shows
evidence for a galactic outflow. This supports the hypothesis that a DLA can be the
gaseous halo of an LBG. 
In addition, we report a redshift coincidence of different objects associated
with metal lines in the same field, separated by 130--161 kpc.
The high detection rate of three correlated structures on a length scale as small
as $\sim150$ kpc in two pairs of lines of sight is intriguing. The absorbers in
each of these are most likely not part of the same gravitationally bound
structure. They more likely represent groups of galaxies.
}


\maketitle

\section{Introduction}
Intervening absorption-line systems found in quasi-stellar-object (quasar, QSO)
spectra play an important role in the observational study of galaxy formation
and evolution. Unlike emission-selected objects, QSO absorption-line systems probe
structures over the full mass range from under-dense regions to massive virialised
dark matter haloes \citep{Rauch1998a,Wolfe2005a}. The most neutral-hydrogen-rich
absorption systems are called damped Ly$\alpha$ absorbers (DLAs), if the neutral
hydrogen column density, $N\left(\ion{H}{i}\right)$, exceeds $2\times10^{20}$
cm$^{-2}$ ($\log\,N\left(\rm{\ion{H}{i}}\right)\geq 20.3$) and sub-DLAs (or sometimes
super-Lyman-limit systems) if the column density in neutral hydrogen is between
$\sim10^{19}$ and $2\times10^{20}$ cm$^{-2}$ \citep[][and ref. therein]{Wolfe2005a}.
DLAs represent the main reservoir of neutral hydrogen in the high-$z$ ($z>2$)
Universe \citep{Wolfe1986a,Peroux2005a, Wolfe2005a}. Since the advent of QSO
absorption-line spectroscopy, over 1000 intervening sub-DLAs and DLAs have been
detected \citep{Prochaska2005a,Noterdaeme2009a,Prochaska2009a}. 

Although their frequency and chemical composition are well known,
the nature of their galaxy counterparts (hereafter called DLA galaxies) has remained
poorly constrained for many years. Various models for DLA galaxies
exist: e.g. rapidly-rotating proto-galactic disks at high redshift
\citep{Prochaska1997a, Prochaska1998a, Wolfe1998a}, low surface brightness
galaxies \citep{Jimenez1999a}, faint and small gas-rich dwarf galaxies
\citep{Tyson1988a, Haehnelt1998a, Okoshi2005a}, compact galaxies
\citep{Nagamine2007a}, and gaseous haloes of high-$z$ LBGs
\citep{Fynbo1999a,Moller2002a,Fynbo2008a,Rafelski2011a}. \citet{Chen2003a} showed that
the majority of low-$z$ sub-DLAs and DLAs are late-type galaxies and only a few are
elliptical or dwarf galaxies. The
conclusion that these low-$z$ absorbers are indeed sub-DLAs or DLAs is drawn
indirectly. Below $z\lesssim1.5$, the Ly$\alpha$ absorption line is below 3000\,\AA,
a spectral range that is inaccessible with ground-based telescopes. \citet{Ellison2006a} showed that
\ion{Mg}{ii} absorbers with $EW_{\rm rest} \left(\rm{\ion{Mg}{ii}\,\lambda2796}\right)\geq1\,\rm\AA$,
so-called strong \ion{Mg}{ii} absorbers, are likely sub-DLAs or DLAs
\citep[see also][]{Rao2000a,Ellison2009a}.

The
successfully identified DLA galaxies allow the study of objects at the very faint end of
the galaxy luminosity function (LF), objects that are usually missed in galaxy
surveys.  During the past 25 years, several deep imaging campaigns have been
carried out, but only with limited success \citep[e.g.][see also \citealt{Bouche2011a}]{Smith1989a,
Djorgovski1996a, LeBrun1997a, Fynbo1999a, Warren2001a, Moller2002a,
Fumagalli2010a}. Up to now, only about 17 sub-DLA and DLA galaxies
have been identified below the redshift of unity \citep[e.g.][and reference therein]
{Peroux2011a} and more than 80 DLA galaxy candidates have been reported to date
\citep[e.g.][]{Rao2011a}. For the majority of these DLA galaxy
candidates there is no spectroscopic confirmation yet.
At high redshift the situation is worse; only 9 intervening
sub-DLAs and DLAs with redshifts between 2 and 3.15 have a confirmed
galaxy counterpart 
\citep[high-$z$;][and references therein]{Djorgovski1996a, Moller2002a,
Christensen2004a, Moller2004a, Weatherley2005a, Fynbo2010a, Fynbo2011a,
Peroux2011b, Bouche2011a, Krogager2012a, Noterdaeme2012a}. In addition to those, a few associated DLAs, 
i.e. $z_{\rm abs} \approx z_{\rm em}$, have detected galaxy counterparts
\citep[e.g.][]{Fynbo1999a, Moller1993a, Ellison2002a, Moller2004a, Adelberger2006a}.

Apart from the faintness of the galaxy counterparts, a further issue
hampering the search for DLA galaxies is the glare of the bright background
QSOs. Counterparts with a small angular distance from the QSO (impact parameter)
are difficult to recover. This may lead to a possible misidentification of
the DLA galaxy and hence an overestimation of its size, as DLAs at small impact parameters
are easily missed.
A different class of background light sources, gamma-ray bursts
(GRBs), provide a complementary approach to the problem. A GRB is a transient
phenomenon, outshining the known Universe in $\gamma$-rays for a fraction of a
second up to hundreds of seconds \citep[e.g.][]{Kouveliotou1993a, Zhang2004a}.
This short-lived emission is followed by an afterglow that can usually be
detected from the X-rays over optical/near-infrared (NIR) wavelengths to the radio for several
weeks \citep[e.g.][]{Racusin2009a,Kann2010a,Chandra2011a}.
An afterglow can outshine the brightest quasars by several
orders of magnitudes, but only for a very short period of time ($\sim 1\,\rm
hr$ -- $\sim1\,\rm day$; e.g.  \citealt{Bloom2009a, Kann2010a}). Similar to QSOs,
GRBs can be found over most of the observable Universe, the most distant
spectroscopically-confirmed 
burst being GRB\,090423, at $z\sim8.3$ \citep{Tanvir2009a, Salvaterra2009a}. In fact, since
most GRBs are associated with the death of massive stars \citep{Hjorth2011a, Woosley2011a},
this should allow us to detect them at higher redshifts than QSOs.
In short, GRB afterglows can be brighter than QSOs but they are ephemeral,
i.e. they vanish within a couple of months. This leaves the line-of-sight clear
and without any interference from a bright object \citep[e.g.][]{Masetti2003a,
Vreeswijk2003a, Jakobsson2004a, Ellison2006b, Prochter2006a, Henriksen2008a, Pollack2009a}.

GRBs are important for DLA studies for another reason. Nearly 
all GRB host galaxies have a DLA (GRB-DLA; \citealt{Fynbo2009a}). The
distribution of their \ion{H}{i} column density, and their metal-line strength distributions
of $\rm{\ion{Si}{ii}}\,\lambda1526$ and $\rm{\ion{C}{iv}}\,\lambda\lambda1548\&1550$
overlap with those of intervening DLAs, but also extend to larger values
\citep{Prochaska2007b, Fynbo2009a}. On average GRB-DLAs have larger metallicities than 
intervening DLAs \citep{Fynbo2006a,Prochaska2007b,Fynbo2008a}.
The difference between in-situ and intervening DLAs lies in the way they probe their host
galaxies \citep{Vreeswijk2004a, Prochaska2007b, Fynbo2008a, Fynbo2009a}. As GRBs are
thought to originate from the collapse of a massive star, GRB-DLAs probe the line of
sight to the location of a massive star, irrespective of the DLA orientation relative to
the host galaxy if the progenitor is located in the DLA. Consequently, GRB-DLAs
are selected by their star-formation rate ($SFR$). In contrast, the properties of an
intervening DLA depend on the geometry of the DLA galaxy and its orientation relative
to the line of sight. These are selected by their covering fraction and hence their
cross-section, $\sigma\left(\rm{\ion{H}{i}}\right)$. A debated question in observational
galaxy formation and evolution is the faint-end slope of the $z\sim3$ galaxy LF. DLA galaxies
probe the faint-end slope, but the in-situ and intervening DLA LFs are only identical
if $\sigma\left(\rm{\ion{H}{i}}\right)\propto SFR$ \citep{Chen2000a, Fynbo2008a}.
The studies of both DLA populations therefore complement each other well.

Currently, seven intervening sub-DLAs and DLAs have been found in six GRB
afterglow spectra, with absorber redshifts ranging between 2.077 and 3.564
(GRBs 050730, 050820A, 050908, 050922C, 060607A, 070721B; \citealt{Chen2005a,
Fox2008a, Piranomonte2008a, Chen2009a, Fynbo2009a, Vergani2009a}).  Although
the number of intervening (sub-)DLAs towards GRBs does not increase the number
of known (sub-)DLAs significantly, the transient nature of the background afterglow simplifies the search for their galaxy
counterparts.  After the afterglow has faded, deep follow-up observations
are usually carried out to find the GRB host galaxy, typically reaching a
limiting magnitude of $\sim27$ mag in the $R$-band \citep[e.g.][]{Hjorth2012a,Malesani2012a}. These
observations are also suitable for the search for DLA galaxies. 

In this paper we present the findings from our search for the photometric
counterparts of intervening sub-DLAs and DLAs in six GRB lines of sight. In Sect.
\ref{sec:reduction}, we introduce our methodology, present the sample selection,
and describe how the data were reduced. The results are then presented in Sect.
\ref{sec:results}, confronted with current models of sub-DLAs and
DLAs, and compared to low-$z$ and high-$z$ DLA galaxies in Sect.~\ref{sec:discussion}.
In Sect.~\ref{sec:conclusion} we draw our conclusions.

Throughout the paper we refer to the Solar abundance compiled in \citet{Asplund2009a}
and adopt $\rm{cm}^{-2}$ as the linear unit of column densities, $N$. We assume
a $\Lambda$CDM cosmology with $H_0 =
71\,\rm{km\,s}^{-1}\,\rm{Mpc}^{-1}$, $\Omega_{\rm m} = 0.27$, and
$\Omega_{\Lambda} = 0.73$ \citep{Larson2011a}.

\section{Data gathering and reduction}\label{sec:reduction}

\subsection{Sample selection and data gathering}\label{sec:selection}

\begin{table}[t!]
\caption{Summary of photometric and spectroscopic data}
\tiny
\begin{tabular}{l c c c c}

\toprule
\multicolumn{4}{c}{\textbf{Photometry}\tablefootmark{a}}\\[1mm]
\multirow{2}{*}{Instrument}	& \multirow{2}{*}{Filter}	& Exposure			& \multirow{2}{*}{Seeing}	& Photo-		\\
							&							& time (s)			&							& metric?		\\
\midrule
\multicolumn{5}{l}{\textbf{GRB 050730: $RA=14^{\rm h}08^{\rm m}17\fs11$ $DEC=-03^\circ46'17\farcs70$ (J2000)}}\\
\midrule
VLT/FORS2				& $R$									& $8\times250 + 24\times300$	& 0\farcs6	& yes	\\
\hst/ACS				& $F775W$								& $6\times1307$					&			&		\\
VLT/ISAAC				& $K_s$									& $32\times60$					& 0\farcs7	& yes	\\
\textbf{\spitzer/IRAC}	& \textbf{$3.6\,\&\,5.8\,\mu\rm m$}		& 7200					&			&		\\
\midrule
\multicolumn{5}{l}{\textbf{GRB 050820A:	$RA=22^{\rm h}29^{\rm m}38\fs114$, $DEC= 19^\circ33'36\farcs61$ (J2000)}}\\
\midrule
Keck/LRIS				& $g$									& 2620							& 0\farcs7	& yes	\\
\hst/ACS				& $F625W$								& 2238							&			&		\\
VLT/FORS2				& $R$									& $4\times500$					& 0\farcs8	& no	\\
\hst/ACS				& $F775W$								& 4404							&			&		\\
\hst/ACS				& $F850LP$								& 14280							&			&		\\
Magellan/PANIC			& $H$									& 8460							& 0\farcs6	& 		\\
VLT/ISAAC				& $K_s$									& $32\times60$					& 0\farcs7	& yes	\\
\textbf{\spitzer/IRAC}	& \textbf{$3.6\,\&\,5.8\,\mu\rm m$}		& 3600							&			&		\\
\midrule
\multicolumn{5}{l}{\textbf{GRB 050908: $RA=01^{\rm h}21^{\rm m}50\fs73$, $DEC=-12^\circ57'17\farcs30$ (J2000)}}	\\
\midrule
VLT/FORS2				& $R$									& $12\times500$					& 0\farcs7	& yes	\\
VLT/ISAAC				& $K_s$									& $32\times60$					& 0\farcs6	& yes	\\
\textbf{\spitzer/IRAC}	& \textbf{$3.6\,\&\,5.8\,\mu\rm m$}		& 7200							&	 		& 		\\
\midrule
\multicolumn{5}{l}{\textbf{GRB 070721B: $RA=02^{\rm h}12^{\rm m}32\fs95$, $DEC=-02^\circ11'40\farcs80$ (J2000)}}\\
\midrule
VLT/FORS2				& $R$									& $24\times235$					& 0\farcs5	& no	\\
\hst/ACS				& $F775W$								& $6\times1307$					& 			&		\\
Magellan/PANIC			& $H$									& 10320							& 0\farcs5	&		\\
VLT/ISAAC				& $K_s$									& $32\times60$					& 1\farcs0	& yes	\\
\bottomrule
\end{tabular}
  \tablefoot{Co-ordinates were taken from \citet{Malesani2012a}.
	Observing conditions of Keck and Magellan data are quoted from \citet{Chen2009a}.
	Column "Photometric?" states if observing conditions were photometric.
	\tablefoottext{a} Date of observations.
	GRB 050730:	VLT/FORS2 - 2006/03/04 - 05/25, \hst/ACS - 2010/06/10, VLT/ISAAC - 2006/03/20, \spitzer/IRAC: 2008/03/11;
	GRB 050820A: 	Keck/LRIS - July 2006, \hst/ACS - $F625W$, $F775W$, $F850LP$: 2005/09/26 and 2006/06/05-11, Magellan/PANIC: August 2007, VLT/FORS2 - 2006/05/24, VLT/ISAAC - 2006/05/20, \spitzer: 2007/12/23;
	GRB 050908:	VLT/FORS2 - 2007/07/21-08/14, VLT/ISAAC - 2007/07/11, \spitzer/IRAC: 2008/08/18;
	GRB 070721B:	VLT/FORS2 - 2007/10/04-11/13, \hst/ACS - 2010/11/13, Magellan/PANIC - August 2007, VLT/ISAAC - 2007/09/22.
} 
\label{tab:log_photometry}
\end{table}

All six GRBs with intervening sub-DLAs and DLAs were a part of deep imaging campaigns 
dedicated to detect their host galaxies. Among these, we selected those that had sufficient
spectroscopic data,
e.g. spectra obtained with different position angles (PAs), or multi-filter data to construct
the spectral energy distribution (SED) of candidates and determine their nature. This
reduced the set to GRBs 050730, 050820A, 050908 and 070721B.

In the following, we briefly describe the photometric and spectroscopic data analysed,
as summarised in Table~\ref{tab:log_photometry}. The observing conditions during Keck
and Magellan observations are described in \citet{Chen2009a}, while 
those of other ground-based observations are summarised in
Table~\ref{tab:log_photometry}. The astrometric uncertainty is $\sim0\farcs3$;
the uncertainty between the optical and NIR astrometry is similar,
allowing us to unambiguously identify objects in the different bands

\subsubsection{GRB\,050730}
Afterglow spectra of GRB\,050730 were acquired with VLT/FORS2 with the 300V grating and with VLT/UVES
with the red and blue arm centred at 3460 \AA \ and 5800 \AA, respectively (see Table~\ref{tab:log_photometry}).
In addition, a further spectrum was acquired with VLT/FORS1 (grating 600V; PA=22.5\fdg), as a part
of the TOUGH (The Optically Unbiased GRB Host) survey \citep{Hjorth2012a}
and three further spectra with VLT/FORS2 (PI: Ellison; Table~\ref{tab:log_photometry}). For the latter, 
the slits  were centred on the afterglow position, and spectra with different slit orientations were obtained 
with the 600B grism.
In doing so, most of the field of view close to the GRB position is covered including several of
the brightest close-by objects. This strategy enables us to localise emission line objects
via triangulation, similar to \citet{Moller2004a},
\citet{Fynbo2010a}, \citet{Fynbo2011a}, \citet{Krogager2012a}, \citet{Noterdaeme2012a}.\footnote{In fact, the galaxy counterparts
of most sub-DLAs and DLAs at $z>2$ were detected via triangulation.}

\citet{Fynbo2009a} reported the serendipitous discovery
of a quasar at the same redshift as the intervening sub-DLA towards GRB\,050730, 17\farcs5
south of the afterglow position. The quasar is blended with a 18.2-mag bright foreground star. In order
to investigate the QSO-sub-DLA connection and to constrain the quasar radiation field, we acquired a
spectrum of the quasar with VLT/X-shooter (Table~\ref{tab:log_photometry}). The field was also the
target of deep imaging campaigns with VLT/FORS2 ($R$-band) and VLT/ISAAC ($K_{\rm s}$-band)
as a part of the TOUGH survey \citep{Hjorth2012a}, \hst/ACS ($F775W$-band, PI: Levan)  and \spitzer/IRAC (3.6 and
$5.8\,\mu\rm m$; Table~\ref{tab:log_photometry}).

\begin{table*}[t!]
%\caption{Summary of spectroscopic data}
\centering
\tiny
\begin{tabular}{l c c c c c c c c}

\toprule
\multicolumn{9}{c}{\textbf{Spectroscopy}\tablefootmark{a}}\\[1mm]
\multirow{2}{*}{Instrument}	& \multirow{2}{*}{Grating}	& Spectral 		& Resolving	& Exposure 	& Slit 		& Position	& \multirow{2}{*}{Seeing}	& Photo-	\\
				&						& range (\AA)		& power		& time (s)	& width		& angle		&				& metric?	\\
\midrule
\multicolumn{5}{l}{\textbf{GRB 050730}}\\
\midrule
VLT/FORS1											& 600V+GG435	& 4430--7370					& 990								& $5\times1316$				& 1\farcs3					& 22\fdg5					& 0\farcs8					& yes\\
\multirow{4}{*}{VLT/FORS2}							& 600B			& 3300--6210					& 780								& $4\times1316$				& 1\farcs3					& -20\fdg0					& 0\farcs9					& yes\\
													& 300V			& 3400--9500					& 440								& $600+1200$				& 1\farcs0					& 0\fdg0					& 1\farcs5 					& no\\
													& 600B			& 3300--6210					& 780								& $5\times1316$				& 1\farcs3 					& 33\fdg5 					& 1\farcs2					& yes\\
													& 600B			& 3300--6210 					& 780								& $4\times1316$				& 1\farcs3					& 95\fdg0					& 1\farcs3					& yes\\
\multirow{2}{*}{VLT/UVES}							& 346 and 580	& 3030--6840\tablefootmark{b}	& \multirow{2}{*}{$\sim40000$}		& \multirow{2}{*}{3000}		& \multirow{2}{*}{1\farcs0}	& \multirow{2}{*}{0\fdg0}	& \multirow{1}{*}{1\farcs8}	& \multirow{1}{*}{no}\\
													& 437 and 860	& 3730--10600\tablefootmark{c}	&									&							&							&							& \multirow{1}{*}{1\farcs1}	& \multirow{1}{*}{yes}\\
\multirow{3}{*}{VLT/X-shooter\tablefootmark{d}}		& UVB			& 3000--5500					& 4000								& $2\times900$				& 1\farcs3					& \multirow{3}{*}{43\fdg7}	& \multirow{3}{*}{0\farcs8} & \multirow{3}{*}{yes}\\
													& VIS			& 5500--10000					& 6700								& $2\times900$				& 1\farcs2					& 			\\
													& NIR			& 10000--25000					& 4000								& $6\times300$				& 1\farcs2					& 			\\
\midrule
\multicolumn{5}{l}{\textbf{GRB 050820A}}\\
\midrule
\multirow{6}{*}{VLT/FORS2}							& 300V			& \multirow{3}{*}{3400--9500}	& \multirow{3}{*}{440}				& $4\times1316$				& 1\farcs3					& -52\fdg0					& 1\farcs2					& yes\\
													& 300V			& 								& 									& $8\times1316$				& 1\farcs3					& 34\fdg0					& 1\farcs3					& yes\\
													& 300V			& 								& 									& $4\times1316$				& 1\farcs3					& 83\fdg0					& 0\farcs7					& yes\\
													& 1028z+OG590	& \multirow{3}{*}{7730--9480}	& \multirow{3}{*}{2560}				& $2\times1316$				& 1\farcs3					& -52\fdg0					& 0\farcs8					& yes\\
													& 1028z+OG590	& 								&									& $4\times1316$				& 1\farcs3					& 34\fdg0					& 1\farcs2					& yes\\
													& 1028z+OG590	& 								& 									& $2\times1316$				& 1\farcs3					& 83\fdg0					& 0\farcs7					& yes\\
\multirow{2}{*}{VLT/UVES}							& 390 and 564	& 3260--6680\tablefootmark{e}	& \multirow{2}{*}{$\sim40000$}		& \multirow{1}{*}{1800+1815}& \multirow{2}{*}{1\farcs0}	& \multirow{2}{*}{0\fdg0}	& \multirow{2}{*}{1\farcs0}	& \multirow{2}{*}{yes}\\
													& 437 and 860	& 3730--10600\tablefootmark{c}	& 									& \multirow{1}{*}{2430}		& 							&							& 							& \\
\midrule
\multicolumn{5}{l}{\textbf{GRB 050908}}	\\
\midrule
Gemini/GMOS-N										& B600+G0305	& 5118--7882					& 844								& $2\times1200$				& 0\farcs75					& 110\fdg0					& 							& \\
\multirow{2}{*}{VLT/FORS1}							& 300V			& 3400--9500					& 440								& $3\times1200$				& 1\farcs0					& 0\fdg0					& 0\farcs6 					& yes\\
													& 600B			& 3300--6210					& 780								& $6\times1345$				& 1\farcs3					& 53\fdg8					& 1\farcs2 					& yes\\
\midrule
\multicolumn{5}{l}{\textbf{GRB 070721B}}\\
\midrule
VLT/FORS1											& 300V			& 3400--9500					& 440								& $5\times2735$				& 1\farcs3					& -118\fdg5					& 1\farcs2					& yes\\
VLT/FORS2											& 300V			& 3400--9500					& 440								& $4\times1800$				& 1\farcs0					& -118\fdg3					& 1\farcs2 					& yes\\
\bottomrule
\end{tabular}
  \tablefoot{
	Observing conditions of the afterglow spectroscopy campaigns, and the VLT/UVES data of GRB
	050730 and of GRB 050820A are taken from \citet{Fynbo2009a}, \citet{Ledoux2009a}, and
	\citet{Vergani2009a}.
	The VLT/UVES and VLT/X-shooter spectrographs consist of two and three arms, respectively, displayed
	in the column "Grating". Observations in the different arms are carried out simultaneously.
	Resolving power of VLT/FORS1+2 and Gemini grisms are given for a slit width of $1''$.
	The stated resolving power of the X-shooter NIR spectrum was decreased during the data reduction
	to increase S/N. The effective resolving power of the X-shooter NIR arm is around ten times
	lower.
	\tablefoottext{a} Date of observations.
	GRB 050730:	VLT/FORS1 600V+GG435 - 2007/02/23-24, VLT/FORS2 600B - $\rm{PA}=20\fdg0$: 2006/06/22, $\rm{PA}=0\fdg0$: 2005/07/31, $\rm{PA}=33\fdg5$: 2006/04/28-05/28, $\rm{PA}=95\fdg0$: 2006/05/28, VLT/UVES: 2005/07/31, VLT/X-shooter: 2010/04/18;
	GRB 050820A: 	VLT/FORS2 300V - $\rm{PA}=-52\fdg0$: 2006/07/22-08/16, $\rm{PA}=34\fdg0$: 2006/08/31-09/22, $\rm{PA}=83\fdg0$: 2006/08/17-09/15, 1028z - $\rm{PA}=-52\fdg0$: 2006/08/16, $\rm{PA}=34\fdg0$: 2006/09/17-22, $\rm{PA}=83\fdg0$: 2006/08/17, VLT/UVES: 2005/08/20;
	GRB 050908:	Gemini/GMOS-N - 2005/09/08, VLT/FORS1 - 300V: 2005/09/08, 600B: 2007/08/16-09/06; 
	GRB 070721B:	VLT/FORS1 - 2007/11/16-12/04, VLT/FORS2 - 2007/07/22.
	\tablefoottext{b} Spectral range of the blue and red arms are 3030--3880 \AA~and 4760--6840 \AA, respectively.
	\tablefoottext{c} Spectral range of the blue and red arms are 3730--4990 \AA~and 6600--10600 \AA, respectively.
	\tablefoottext{d} The target of the X-shooter observation was QSO J1408-0346.
	\tablefoottext{e} Spectral range of the blue and red arms are 3260--4450 \AA~and 4580--6680 \AA, respectively.
} 
%\label{tab:log_photometry_2}
\hfill\center{Tab. \ref{tab:log_photometry} --- continued}
\end{table*}

\subsubsection{GRB\,050820A}\label{ref:paper_intervening_050820a}
Similar to GRB\,050730, GRB\,050820A was a target of an extensive spectroscopic campaign.
After the afterglow
faded, several spectra were acquired with VLT/FORS2 by centring the slit on the afterglow
position and using three different slit orientations, each one observed with two different
gratings (300V and 1028z; PI: Ellison), see Table~\ref{tab:log_photometry}. Slits
were oriented to cover most of the area around the afterglow position and
several of the brightest objects close to the afterglow position. The total integration
time of the data acquired with the 1028z grism is only half of the time spent on 300V, and the 1028z spectral range
does not extend the range of the 300V data substantially, but its resolution is much higher than that of the 300V data.
Hence, we do not use the 1028z data
in our analysis. In addition, the field was observed in nine filters extending
from 473.1 nm ($g'$ band) to $5.8\,\mu\rm m$ with VLT/FORS2 and VLT/ISAAC
as a part of the TOUGH survey, \hst/ACS,
KECK/LRIS, Magellan/PANIC, and \spitzer/IRAC (Table~\ref{tab:log_photometry}).

\subsubsection{GRB\,050908}

For this burst, the amount of available data is small. An afterglow spectrum was obtained with the
300V grating (slit width of 1\farcs0 and $\rm{PA}=0\fdg0$) utilising VLT/FORS1, and with Gemini/GMOS-N
utilising the B600 grating (slit width of 0\farcs75 and $\rm{PA}=110\fdg0$; 
Table~\ref{tab:log_photometry}). In addition to these data, we included a spectrum obtained with
VLT/FORS1 (600B) and deep $R$- and $K_{\rm s}$-band images acquired with VLT/FORS2 and VLT/ISAAC
as a part of the TOUGH survey,
respectively, and deep \spitzer/IRAC images acquired at 3.6 and $5.8\,\mu\rm m$
(Table~\ref{tab:log_photometry}).

\subsubsection{GRB\,070721B}
We use the afterglow spectrum (PA=-118\fdg3) acquired with
VLT/FORS2 (Table~\ref{tab:log_photometry}), and a spectrum obtained with VLT/FORS1 (PA=-118\fdg5) several months
after the GRB faded as a part of the TOUGH campaign. In addition, several deep imaging campaigns targeted this field.
We use the VLT/FORS2 $R$-band and VLT/ISAAC $K_{\rm s}$-band data acquired as a part of the TOUGH campaign,
\hst/ACS F775W-band data, and Magellan/PANIC $H$-band data.

\subsection{Data reduction}\label{subsec:reduction}

\subsubsection{Imaging}\label{sec:imaging}

\paragraph{Ground-based data --} VLT/FORS and Keck/LRIS data were reduced in a standard
way, VLT/FORS data with \iraf \ and Keck/LRIS data with \texttt{IDL} (\citealt{Tody1986a};
see \citealt{Malesani2012a} for more details on the VLT data reduction,
and \citealt{Perley2012a} on the Keck data reduction).
The NIR data acquired with VLT/ISAAC were reduced in a standard way
with the \texttt{jitter} routine in \texttt{eclipse} \citep{Devillard1997a}.
The reduction of the Magellan/PANIC data is presented in \citet{Chen2009a}.

\paragraph{\hst \ data --}

\hst/ACS images of GRB\,050730 and 070721B consist of 6 dithered
exposures in the $F775W$ filter. Individual exposures, after standard
``on-the-fly'' processing, were retrieved from the STScI archive.\footnote{http://archive.stsci.edu}
These were subsequently cleaned for bias striping, introduced due to the replacement electronics after Servicing Mission 4 (May 11-24, 2009),
and then drizzled using the {\tt multidrizzle} software into final science
images. At this stage we adopt a pixel scale of 0\farcs033,
approximately 2/3 of the native pixel scale.
The data reduction of the GRB\,050820A \hst \ data is presented in
\citet{Chen2009a}. The pixel scale of these drizzled images is 0\farcs05
in contrast to that of the images of GRBs 050730 and 070721B.

\paragraph{\spitzer \ Data}

GRBs\,050730, 050820A, and 050908 were observed with the Infrared Array Camera aboard \spitzer~
\citep[IRAC;][]{Fazio2004a} in the bandpasses centred at 3.6 and $5.8\,\mu\rm m$.
We downloaded the processed PBCD data from the \spitzer~Heritage Archive
and followed the basic procedures for aperture photometry in the IRAC
Instrument Handbook,\footnote{http://http://sha.ipac.caltech.edu/applications/Spitzer/SHA/}
using an extraction radius of 2 native pixels (2\farcs4, or 4 pixels in
the default archive resampling).

\paragraph{Photometry --}

In order to measure the total flux, we applied aperture photometry, utilising Source
Extractor \citep{Bertin1996a}, and applied an aperture correction 
assuming a stellar point-spread function (PSF; for the aperture correction of the \hst \
images see \citealt{Sirianni2005a}).

The quasar in the field of GRB\,050730 is strongly blended with a foreground star at a
projected distance of 1\farcs{2}.
To perform reliable photometry, the
contaminating star was subtracted from the VLT/FORS1 images. A model of the
FORS1 PSF was constructed from $\sim$60 field stars using Daophot as implemented
in \iraf. This PSF model was
then fitted simultaneously to the star and the QSO. To test the quality of the fit,
we subsequently subtracted the stellar contribution from the input image. 
Photometry was then performed in the final
star-subtracted image.

Instrumental magnitudes obtained from optical ground-based data were calibrated against
2--3 photometric standard stars \citep{Landolt1992a, Stetson2000a}. NIR magnitudes
were calibrated with more than three 2MASS stars and \hst \ magnitudes against zeropoints,
computed from their FITS headers.

\subsubsection{Spectroscopy}

\paragraph{VLT/FORS1 and FORS2, Gemini/GMOS and VLT/UVES data --}

FORS and Gemini data were reduced in a standard way with \texttt{IRAF} \citep[for more details
see][]{Fynbo2009a}. The data reduction of the UVES data is described in \citet{Ledoux2009a}.

\paragraph{VLT/X-shooter data --}

\begin{figure*}
\centering
\includegraphics[width=1.0\textwidth, angle=0]{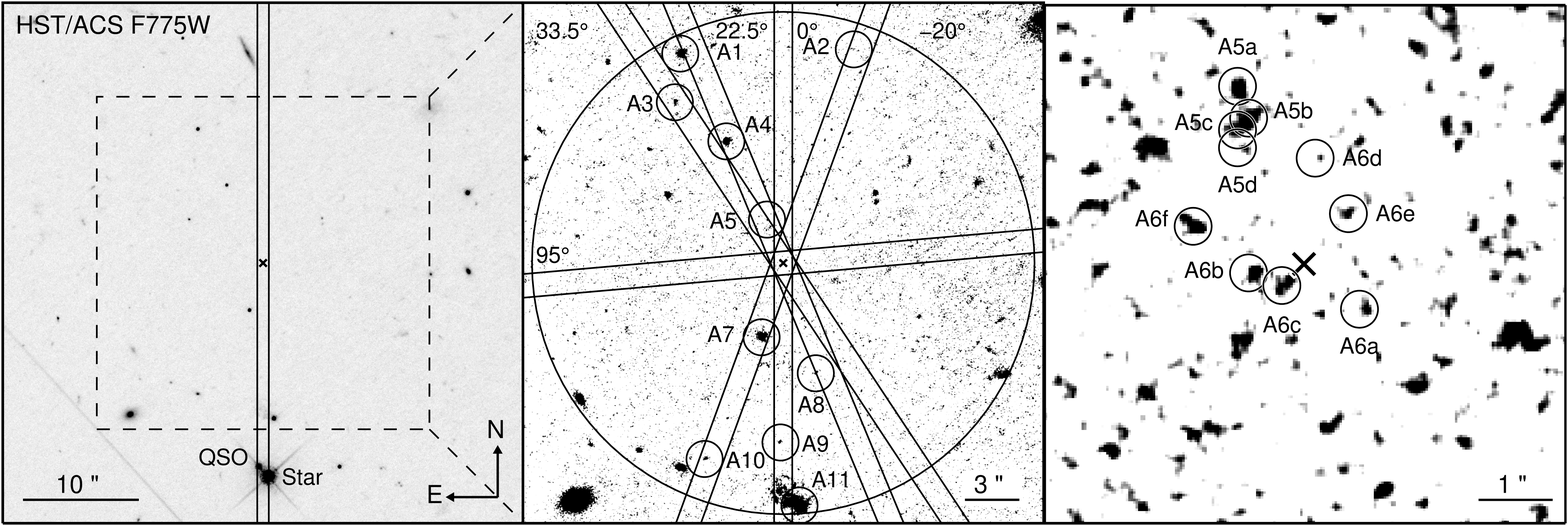}
\caption{\textit{Left panel}: The field of view of GRB\,050730 ($z=3.969$) observed in $F775W$ with \hst/ACS.
	\textit{Middle panel}: Zoom-in to the inner 28\farcs8. The different
	slit orientations from Table~\ref{tab:log_photometry} are over-plotted.
	\textit{Right panel}: Zoom-in to the inner 6\arcsec; the panel was
	smoothed by a boxcar filter (width: $0\farcs132 = 4\,\rm pixel$).
	The afterglow position is marked by an '$\times$' in all panels. The 100-kpc
	radius (shown in the middle panel) represents the assumed maximum impact parameter of the absorber
	at $z=3.56$. Objects are labelled if they lie within the circle and if
	they are covered by any of the slits. Their magnitudes are summarised
	in Table~\ref{tab:candidates}.
	}
\label{fig:fov_050730}
\end{figure*}

The QSO in the field of GRB\,050730 was observed with X-shooter. These data were reduced with the
X-shooter pipeline v1.2.2.\footnote{http://www.eso.org/sci/software/pipelines/}
The 1-dimensional spectrum of the QSO ($R=20.76\,\rm mag$; corrected for Galactic extinction) in the field
of GRB\,050730 could not be extracted with the X-shooter pipeline, 
because of being blended with the 18.2-mag
bright foreground star 1\farcs2 SW mentioned above (see Fig.~\ref{fig:fov_050730}). 

\begin{table*}
\caption{Properties of the intervening sub-DLAs and DLAs}
\centering
\begin{tabular}{lccccccl}
\toprule
\multirow{2}*{GRB}	&\multirow{2}*{$z_{\rm GRB}$}	&\multirow{2}*{$z_{\rm abs}$}	&$\log\,N\left(\rm{\ion{H}{i}}\right)$	&\multirow{2}*{$\left[\rm{Si}/\rm{H}\right]$}	&$EW_{\rm rest}\left(\rm{\ion{Si}{ii}}\,\lambda1526\right)$	& \multirow{2}*{Detected lines}		& \multirow{2}*{References}	\\
			&				&				&					& 							&$\left(\rm{\AA}\right)$					&					& \\
\midrule
			& 				& 3.56439			& $20.3\pm0.1$				&$<-1.3$					& $<1.01$\tablefootmark{a}		& \ion{Al}{ii}, \ion{Fe}{ii}, \ion{Si}{ii}, \ion{Si}{ii}$^*$, \ion{C}{iv}, \ion{Si}{iv}	& \\
\multirow{-2}*{050730}	& \multirow{-2}*{3.969}		& 3.02209			& $19.9\pm0.1$				&$-1.5\pm0.2$\tablefootmark{b}			& \dots					& \ion{Al}{ii}, \ion{Fe}{ii}, \ion{Si}{ii}						& \multirow{-2}{*}{1, 2, 3}\\[1mm]
 
050820A			& 2.615				& 2.3598\tablefootmark{c}	& $20.1\pm0.2$				&$-1.5\pm0.2$					& $0.17\pm0.01$\tablefootmark{d}	& \ion{Fe}{ii}, \ion{Si}{ii}, \ion{C}{iv}, \ion{Mg}{ii}					& 4\\[1mm]

050908			& 3.3467			& 2.6208			& $20.8\pm0.1$\tablefootmark{d, e}	& $>-1.40$\tablefootmark{d}			& $2.24\pm0.06$				& \ion{Al}{ii}, \ion{Fe}{ii}, \ion{Si}{ii}, \ion{C}{iv}					& 2\\[1mm]

070721B			& 3.6298			& 3.0939\tablefootmark{g}	& $20.1\pm0.3$\tablefootmark{d, e}	& $>-0.66$\tablefootmark{d}			& $2.35\pm0.69$\tablefootmark{d}	& \ion{Al}{ii}, \ion{Fe}{ii}, \ion{Si}{ii}, \ion{C}{iv}					& 2\\
\bottomrule
\end{tabular}
\tablefoot{
	\tablefoottext{a} This line is blended. The $EW$ listed is the value of the blend.
	\tablefoottext{b} Without ionisation correction.
	\tablefoottext{c} \citet{Vergani2009a} reported $EW_{\rm rest}\left(\rm{\ion{Mg}{ii}}\,\lambda2796\right)<0.42\,\rm{\AA}$ with a most likely value of 0.31\,\AA.
	\tablefoottext{d} This work.
	\tablefoottext{e} \citet{Fynbo2009a} suggested that the absorber is a sub-DLA.
	\tablefoottext{f} Derived in the optically thin limit.
	\tablefoottext{g} \citet{Chen2009a} and \citet{Fynbo2009a} proposed that a galaxy 0\farcs9 SE of the afterglow position
	is the galaxy counterpart to the intervening absorber.}
  \tablebib{
  (1) \citet{Chen2005a};
  (2) \citet{Fynbo2009a};
  (3) \citet{Starling2005a};
  (4) \citet{Vergani2009a}
}
\label{tab:dla_sample}
\end{table*}

\begin{table*}
\caption{Properties of selected \ion{Mg}{ii} absorbers}
\centering
\begin{tabular}{lcccccll}
\toprule
\multirow{2}*{GRB}		&\multirow{2}*{$z_{\rm GRB}$}	&\multirow{2}*{$z_{\rm abs}$}	& $EW_{\rm rest}\left(\rm{\ion{Mg}{ii}}\,\lambda2796\right)$	& \multirow{2}*{Detected lines}			& \multirow{2}*{References}	\\
				&				&				& $\left(\rm{\AA}\right)$					&						& 				\\
\midrule
				&				& 2.25313			& $<0.78\,(0.65)^\dagger$					& \ion{Fe}{ii}, \ion{Mg}{i}, \ion{Mg}{ii}	& 1, 2		\\
\multirow{-2}*{050730}		&\multirow{-2}*{3.969}		& 1.7731\tablefootmark{a}	& $0.93\pm0.03^\dagger$						& \ion{Fe}{ii}, \ion{Mg}{i}, \ion{Mg}{ii}	& 1, 2	\\[1mm]

				& 				& 1.4288			& $1.32\pm0.02$							& \ion{Fe}{ii}, \ion{Mg}{i}, \ion{Mg}{ii} 	& 1		\\
\multirow{-2}*{050820A}		&\multirow{-2}*{2.6147}		& 0.6915			& $2.87\pm0.01^\dagger$						& \ion{Ca}{ii}, \ion{Fe}{ii}, \ion{Mg}{ii}	& 1		\\[1mm]

\multirow{-1}*{050908}		& \multirow{-1}*{3.3467}	& 1.5481			& $0.82\pm0.07^\dagger$						& \ion{Fe}{ii}, \ion{Mg}{ii}			& 3		\\
\bottomrule
\end{tabular}
\tablefoot{We only summarize the properties of those \ion{Mg}{ii} absorbers that most
	likely have an intervening sub-DLA or DLA. Following the typical approach in the literature,
	we call absorbers with
	$EW_{\rm rest}\left(\rm{\ion{Mg}{ii}}\,\lambda2796\right)>1.0\,\rm \AA$
	strong, with $0.3\,\rm{\AA}<EW_{\rm rest}\left(\rm{\ion{Mg}{ii}}\,\lambda2796\right)<1.0\,\rm \AA$
	intermediate and with $EW_{\rm rest}\left(\rm{\ion{Mg}{ii}}\,\lambda2796\right)<0.3\,\rm{\AA}$ weak systems. Absorbers are marked by $^\dagger$ if the doublet is saturated.
	\tablefoottext{a} \citet{Chen2005a} identified two velocity components that are separated by
			$57\,\rm{km}\,\rm{s}^{-1}$.
    }
  \tablebib{
  (1) \citet{Vergani2009a};
  (2) \citet{Prochaska2007a};
  (3) \citet{Fynbo2009a}
}
\label{tab:metal_sample}
\end{table*}

To estimate the flux of the quasar over the entire wavelength range of the X-shooter spectrograph,
we proceeded in the following way.
For each arm, we first extracted a profile of the superposed
spectral PSFs at several wavelengths. Each profile was best described by two Gaussians, representing the QSO
and the star. The centre of their peaks and the widths did not change with wavelength, allowing us to fix
these parameters to their mean values for each arm of the X-shooter spectrograph. Then, we extracted
the profile for every wavelength and fit it with the aforementioned model, using the routine \texttt{mpfit}
by \citet{Markwardt2009a}, in \texttt{IDL}. To identify outliers, a 9-point median profile was fitted for every
wavelength, i.e. four points redward and four points blueward of the considered wavelength. Data points 
were rejected if they deviated by more than $3\sigma$ from the median spectral PSF of the quasar and the star.

\begin{table*}[t!]
\caption{Galaxy counterpart candidates of the intervening absorption line systems}
\centering
\tiny
% \begin{tabular}{l l c@{\hspace{2mm}}c@{\hspace{2mm}}c@{\hspace{2mm}}c@{\hspace{2mm}}c@{\hspace{2mm}}c@{\hspace{2mm}}c@{\hspace{2mm}}c@{\hspace{2mm}}c}
\begin{tabular}{l l ccc ccc cc cc}
\toprule
\multicolumn{10}{l}{\textbf{GRB 050730} (GRB: $z= 3.969$, DLA: $z= 3.56439$, sub-DLA: $z= 3.02209$, \ion{Mg}{ii} absorbers: $z=2.25313$, 1.7731)}\\ 
 Candidate  & Remark & $\theta\,(\arcsec)$ & $R$ (mag)   & $F775W$ (mag)  & $K_s$ (mag)& $3.6\,\mu\rm m$ (mag)& $5.8\,\mu\rm m$ (mag)\\                  
\midrule                                    
\textbf{A1}				&star		&12.8	&$23.46 \pm0.03	$&$23.08\pm0.01	$&$22.51\pm0.25	$&$23.30\pm0.20$\\
A2					&		&12.3	&$		$&$28.50\pm0.26	$&$		$\\						
A3					&		&10.6	&$26.53 \pm0.12	$&$26.60\pm0.09	$&$		$\\			
\textbf{A4}				&star		&7.3	&$23.88 \pm0.03	$&$23.71\pm0.02	$&$		$\\			
A5a					&		&2.5	&$27.20 \pm0.27	$&$27.98\pm0.18	$&$		$\\			
A5b\tablefootmark{a}			&		&2.1	&$		$&$28.16\pm0.19	$&$		$\\						
A5c\tablefootmark{a}			&		&2.0	&$		$&$28.43\pm0.22	$&$		$\\						
A5d					&		&1.8	&$		$&$28.84\pm0.30	$&$		$\\						
A6a					&		&1.0	&$		$&$28.64\pm0.26	$&$		$\\						
A6b					&		&0.7	&$		$&$28.37\pm0.24	$&$		$\\						
A6c					&		&0.4	&$		$&$28.11\pm0.19	$&$		$\\						
A6d					&		&1.4	&$		$&$28.64\pm0.26	$&$		$\\						
A6e					&		&0.9	&$		$&$28.64\pm0.26	$&$		$\\						
A6f					&		&1.5	&$		$&$28.29\pm0.20	$&$		$\\						
\textbf{A7}				&star		&4.2	&$23.38 \pm 0.03$&$22.84\pm0.01	$&$22.28\pm0.15	$&$22.78\pm0.15$\\
A8					&		&6.3	&$		$&$27.49\pm0.12	$&$		$\\						
A9					&		&9.8	&$		$&$27.38\pm0.12	$&$		$\\						
A10					&		&11.5	&$		$&$27.10\pm0.13	$&$		$\\						
\textbf{A11}\tablefootmark{b}		&$z\sim0.16\rm{?;}\,z<2.81$	&13.3	&$		$&$		$&$		$& & \\
\textbf{QSO}\tablefootmark{c}		&$z=3.022$	&17.5	&$20.87 \pm 0.05$&$\sim	20.58	$&$		$\\					
\textbf{Star}\tablefootmark{c}		&star		&18.3	&$18.20 \pm 0.04$&$		$&$		$\\						
\midrule                                    
\multicolumn{10}{l}{\textbf{GRB 050820A} (GRB: $z= 2.615$, DLA: $z= 2.3598$, \ion{Mg}{ii} absorbers: $z=1.4288$, 0.6915)}\\
 Candidate  					&  Remark &$\theta\,(\arcsec)$	 		&  $g$ (mag)  	&  $F625W$ (mag)  &   $R$ (mag)   &  $F775W$ (mag)  &  $F850LP$ (mag)  &  $H$ (mag)  &  $K_s$ (mag)\\   
\midrule                                    
B1\tablefootmark{d}   			&  								& 11.8 	&$    			$&$25.18\pm0.04	$&$	    		$&$24.87\pm0.03$&$24.90\pm0.06	$&$    			$&$    			$\\
\textbf{B2}  					& $z=0.693$ 					& 12.3 	&$ 25.41\pm0.07	$&$23.59\pm0.03	$&$23.20\pm0.03 $&$22.35\pm0.01$&$21.81\pm0.05	$&$20.81\pm0.10	$&$20.47\pm 0.05$\\
\textbf{B3}  					& $z=0.67^{+0.02} _{-0.07}$ 	& 5.2 	&$ 25.24\pm0.06 $&$23.90\pm0.04	$&$23.65\pm0.04 $&$23.01\pm0.02$&$22.57\pm0.05	$&$21.41\pm0.14	$& $21.05\pm 0.07$	\\
\textbf{B4} 					& star 							& 2.8 	&$ 23.39\pm0.05 $&$22.11\pm0.01	$&$21.97\pm0.03 $&$21.68\pm0.01$&$21.34\pm0.01	$&$20.94\pm0.11	$&$21.19\pm 0.07$	\\
\textbf{B5} 					& $z=0.9\pm0.1$ 				& 3.7 	&$ 26.68\pm0.14 $&$26.36\pm0.09	$&$26.48\pm0.26 $&$25.58\pm0.06$&$25.36\pm0.06	$&$    			$&				\\
\textbf{B6} 					& $z=1.5^{+0.2} _{-0.3}$ 		& 4.9 	&$25.23\pm0.06	$&$24.94\pm0.09	$&$24.80\pm0.09	$&$24.32\pm0.04$&$23.67\pm0.05	$&$22.28\pm0.21	$&$22.20\pm0.15	$\\
\textbf{B7}\tablefootmark{a}	& $z=2.615$ 					& 1.6 	&$ 25.75\pm0.08	$&$25.99\pm0.07	$&$25.66\pm0.10	$&$25.90\pm0.06$&$25.83\pm0.08	$&$    			$&    			\\
\textbf{B8N}\tablefootmark{a} 	& \multirow{2}{*}{$z=2.615$} 	& 0.3	&$				$&$26.16\pm0.08	$&$    			$&$25.83\pm0.08$&$26.18\pm0.06	$&$    			$&    			\\
\textbf{B8S}\tablefootmark{a} 	&  								& 0.1 	&$    			$&$26.19\pm0.08	$&$				$&$25.72\pm0.07$&$25.82\pm0.05	$&$    			$&    			\\
\textbf{B9}  					& $z=0.428$ 					& 3.4 	&$ 25.49\pm0.07 $&$24.94\pm0.05	$&$24.45\pm0.07	$&$24.76\pm0.03$&$24.57\pm0.04	$&$    			$&    			\\
\textbf{B10}  					& star 							& 3.6 	&$ 22.11\pm0.05 $&$20.58\pm0.00	$&$20.31\pm0.03	$&$19.69\pm0.00$&$19.16\pm0.00	$&$18.60\pm0.05	$&$18.81\pm0.04$		\\
\midrule                                    
Candidate  						&  	 							& & $3.6\,\mu\rm m$ (mag)& $5.8\,\mu\rm m$ (mag)\\   
\midrule                                    
B1\tablefootmark{d}   			&  								& &$23.45\pm0.30$					\\
\textbf{B2}  					& 								& &$20.59\pm0.05$ &$21.40\pm0.20$	\\
\textbf{B3}  					&								& &$20.94\pm0.06$ & $21.04\pm0.18$	\\
\textbf{B4} 					& 								& &$21.65\pm0.10$ & $21.68\pm0.32$	\\
\textbf{B5} 					& 								& 									\\
\textbf{B6} 					& 								& &$21.81\pm0.10$					\\
\textbf{B7}\tablefootmark{a}	& 								& 									\\
\textbf{B8N}\tablefootmark{a} 	& 								& 									\\
\textbf{B8S}\tablefootmark{a} 	& 								& 									\\
\textbf{B9}  					& 								& 									\\
\textbf{B10}  					& 								& &$19.51\pm0.03$ & $20.29\pm0.09$	\\
\bottomrule                        
 \end{tabular}
\tablefoot{Observed magnitudes are given in the AB system (not corrected for Galactic extinction). VLT and Magellan Vega-magnitudes
    were converted into the AB system by adding: VLT/$R$ 0.21 mag \citep{Blanton2007a},
    VLT/$K_s$ 1.895 mag (ESO), and Magellan/$H$ 1.34 mag. The selective Galactic extinctions, $E(B-V)$,
    are 0.051 and 0.044 mag, for GRBs 050730 and 050820A, respectively.
    Impact parameters were derived from the \hst~images, if available, otherwise from $R$-band images.
    The uncertainty in the impact parameter is $0\farcs3-0\farcs4$. Spectra were extracted for boldfaced
    objects. Based on the spectrum, the SED or the shape and colour of an object, we report in the column
    "remark" if it is a star or at which redshift the galaxy is. The $3\sigma$ limiting magnitudes are:
    GRB 050730: $R=27.1$, $F775W=28.6$, $K_s=22.8$, $3.5\,\mu\rm{m}=23.5$, $5.8\,\mu\rm{m}=21.5$ mag,
    GRB 050820A: $g'=27.6$, $F625W=27.8$, $R=26.1$, $F775W=28.1$, $F850LP=28.0$, $H=23.0$, $K_s=23.0$, $3.5\,\mu\rm{m}=23.5$, $5.8\,\mu\rm{m}=21.7 mag$,
%     GRB 050908: $R=27.4$, $K_s=23.0$, and
%     GRB 070721B: $R=27.5$, $F775W=28.7$, $H=23.1$, $K_s=23.2$.
   \tablefoottext{a} {Blended object. The magnitude is only an estimate. The spectrum can be the
    integrated spectrum of the blended objects.}
   \tablefoottext{b}{The object consists of at least two galaxies and is detected in all filters.
		We do not report any magnitude of the individual objects or the compound in this work. We detect
		the continuum down to 4630 \AA. If this break is caused by Ly$\alpha$ absorption, the redshift is $z=2.81$, or if
		this is the redshifted 4000\,\AA~feature, the redshift is $z=0.16$.}
   \tablefoottext{c} {The $R$-band brightness of the quasar and the star were obtained from an afterglow image
    taken on 23 February 2007 using PSF photometry (Sect. \ref{sec:imaging}). The blended star is
    saturated in the \hst~image. The PSF was subtracted  with Tiny Tim \citep{Krist1993a} assuming
    a $FWHM$ of $5''$. While the wings of the
    star were partly removed, the residuals in the core are quite strong. The measurement of the QSO magnitude
    is only an estimate.}
   \tablefoottext{d}{The compound is a blend of several individual objects. We only report the brightness
   	in the \hst~images of the brightest object.}
}
\label{tab:candidates}
\end{table*}

\begin{table*}
\centering
\tiny
% \begin{tabular}{l l c@{\hspace{2mm}}c@{\hspace{2mm}}c@{\hspace{2mm}}c@{\hspace{2mm}}c@{\hspace{2mm}}c@{\hspace{2mm}}c@{\hspace{2mm}}c@{\hspace{2mm}}c}
\begin{tabular}{l l ccc ccc}
\toprule
% \multicolumn{10}{l}{\textbf{GRB 050730} (GRB: $z= 3.969$, sub-DLA: $z= 3.02209$, DLA: $z= 3.56439$, weak \ion{Mg}{ii} absorber: $2.25313$, strong \ion{Mg}{ii} absorber: $1.7731$)asdfasdf}\\ 
\multicolumn{8}{l}{\textbf{GRB 050908} (DLA: $z= 3.3467$, DLA: $z= 2.6208$, \ion{Mg}{ii} absorber: $z=1.5481$)}\\
 Candidate   						& Remark & $\theta\,(\arcsec)$ & $R$ (mag)  & $K_s$ (mag) & $3.6\,\mu\rm m$ (mag)& $5.8\,\mu\rm m$ (mag)&\\
\midrule                        
\textbf{C1 } 			& $z\approx2.71$& 12.6 &$ 27.10 \pm 0.20 $& &\\            
C2   					& 				& 11.5 &$ 24.96 \pm 0.04 $& &$23.25\pm0.19$\\
C3   					& 				& 6.6 &$ 25.97 \pm 0.09 $& &\\            
C4\tablefootmark{a} 	& 	 			& 12.4 &$ 26.17 \pm 0.09 $& &\\            
C5\tablefootmark{a,b}	&  				& 3.7 &$ 25.05 \pm 0.04 $ && $23.65\pm0.23$&$22.44\pm0.44$\\
C6\tablefootmark{a,b} 	&  				& 2.8 &$ 25.39 \pm 0.05 $& &\\            
C7\tablefootmark{b}		& 				& 1.1 &$ 26.98 \pm 0.18 $& &\\            
C8						&  				& 5.5 &$ 26.59 \pm 0.21 $& &\\            
C9		   				& 				& 2.5 &$ 25.96 \pm 0.12 $& &$\sim25.0\pm0.5$&\\            
\midrule                        
\multicolumn{8}{l}{\textbf{GRB 070721B} (GRB: $z= 3.6298$, DLA: $z= 3.0939$)}\\                       
 Candidate   & Remark & $\theta\,(\arcsec)$ & $R$ (mag)  & $F775W$ (mag)  & $H$ (mag)  & $K_s$ (mag)\\   
\midrule                        
D1   							&  									& 12.8 &$    				$&$ 27.40 \pm 0.12 $&$    $&$    $\\
\textbf{D2a}\tablefootmark{a} 	& \multirow{2}{*}{$z\approx2.64$} 	& 3.0 &$    				$&$ 27.34 \pm 0.12 $&$    $&$    $\\
\textbf{D2b}\tablefootmark{a} 	& 				 					& 2.8 &$   				 	$&$ 28.23 \pm 0.22 $&$    $&$    $\\
\textbf{D3 }  					& $z=3.096$			 				& 1.0 &$ 24.48 \pm 0.02 	$&$ 24.41 \pm 0.02 $&$ 23.40 \pm 0.07 $&$ 23.56 \pm 0.23 $\\
\textbf{D4 } 					& $z=3.631$ 						& 0.1 &$    				$&$ 27.31 \pm 0.15 $&$    $&$    $\\
D5  							&  									& 6.9 &$ 26.82 \pm 0.18 	$&$ 26.44 \pm 0.10 $&$    $&$    $\\
\textbf{D6a}\tablefootmark{a} 	& \multirow{2}{*}{$z\approx2.64$} 	& 11.7 &$    				$&$ 25.05 \pm 0.04 $&$    $&$    $\\
\textbf{D6b}\tablefootmark{a} 	&  									& 12.2 &$    				$&$ 26.90 \pm 0.09 $&$    $&$    $\\
\textbf{D7}  					& $z=3.615$ 						& 20.7 &$ 23.77 \pm 0.01	$&$ 23.74 \pm 0.03 $&$ 24.02 \pm 0.11 $&$ 23.20 \pm 0.17 $\\
\bottomrule
\end{tabular}
\tablefoot{The selective Galactic extinctions, $E(B-V)$, are 0.025 and 0.031 mag for GRBs 050908 and 070721B, respectively.
    Spectra were extracted for boldfaced objects. Based on the spectrum or the SED of an object, we show in the column
    "remark" if it is a star or at which redshift the galaxy is. The $3\sigma$ limiting magnitudes are:
    GRB 050908: $R=27.4$, $K_s=23.0$, $3.6\,\mu\rm m= 23.8$, $5.8\,\mu\rm m= 22.3$ mag and
    GRB 070721B: $R=27.5$, $F775W=28.7$, $H=23.1$, $K_s=23.2$ mag.
    \tablefoottext{a} {Blended object. The magnitude is only an estimate. The spectrum is the
    integrated spectrum of the blended objects.}
    \tablefoottext{b} {We do not attempt photometry on objects which are strongly blended in
    the \spitzer~data in this work, with the exception of objects C5, C6, and C7, which are sufficiently closely
    blended that we report the (summed) photometry of all three objects as a group.}
}
\hfill\center{Tab. \ref{tab:candidates} --- continued}
\end{table*}

The QSO was barely visible in the NIR spectrum. We therefore determined the centre of peak and the $FHWM$
of the star and set the QSO centre of peak to its expected position. Furthermore, we
reduced the spectral resolution to increase the signal-to-noise ratio (S/N) by rebinning the spectrum.
The uncertainties in the $FWHM$s and the centre of peaks varied between 0.06 and 0.39 px and 0.3 and 0.18 px
for each arm (pixel scale: UVB and VIS 0\farcs16/px, NIR 0\farcs21/px), respectively, small enough to be neglected.

We compared the quality of our method with the MCS deconvolution technique developed by
Magain, Courbin, \& Sohy \citep{Magain1998a} used in \citet[][see also \citealt{Courbin2000a}]{Letawe2008a}.
This method requires an unblended star to be observed with the identical instrument setup.
Because of the short slit length, no such object was covered by the slit. For a rough estimate,
we used the observed standard star as reference, although it was observed with a slit width of
5\arcsec. Within the errors both methods give the same results. 

To flux calibrate the spectrum, we reduced the data of the standard star GD50 with the
X-shooter pipeline. The 1D spectrum was extracted using the routine \texttt{apall}
in optimal extraction mode. 
The spectrum was then divided by the
corresponding reference spectrum from the \texttt{CALSPEC HST} database \citep{Bohlin2004a}
to deduce the response function. In addition, we corrected the flux-calibrated standard star spectrum
for undulations by smoothing the ratio between the observed flux-calibrated and expected spectra with
a Hamming filter (window size: 40\AA~in the UVB and VIS arm) in regions that were not affected by strong telluric lines or
stellar absorption features. We then applied the corrected response function to the QSO and used
the acquisition image to secure the absolute flux calibration. Finally, we followed \citet{Cardelli1989a}
to correct the QSO for Galactic dust attenuation ($E\left(B-V\right)=0.048\,\rm mag$). No attempt
was made to correct for telluric absorption lines due to the lack of a suitable telluric 
standard star observed the same night. This has no implications on our 
analysis.

\subsection{SED fitting}\label{sec:sed_fitting}

The photometric SEDs of the galaxies were modeled within \texttt{Le Phare} \citep{Arnouts1999a,
Ilbert2006a}.\footnote{http://www.cfht.hawaii.edu/~arnouts/LEPHARE} Here, we used a grid of galaxy templates
based on \citet{Bruzual2003a} stellar population synthesis models with a Chabrier IMF \citep{Chabrier2000a}
and a Calzetti \citep{Calzetti2000a} dust attenuation curve. In those cases, where prior information about
the redshift was available from spectroscopy, we fixed this parameter to the respective value. For a
description of the galaxy templates, physical parameters of the galaxy fitting and their error estimation we
refer to \citet{Kruehler2011a}. To account for zeropoint offsets in the cross calibration and absolute flux
scale, an systematic error contribution of 0.05 mag was added in quadrature to the uncertainty introduced by
photon noise.

\section{Results}\label{sec:results}

Afterglow spectroscopy allowed us to identify seven intervening sub-DLAs and DLAs in six GRB
fields. For four of these fields, we had sufficient data to elucidate the nature of several
galaxy counterpart candidates. In the following, we present how candidates were selected, and
present the properties of the intervening sub-DLAs and DLAs towards GRBs 050730, 050820A,
050908 and 070721B and our findings on the most likely galaxy counterpart for each field.
A summary of the properties of the intervening sub-DLAs and DLAs
is presented in Tables \ref{tab:dla_sample} and \ref{tab:metal_sample}. Most of the shown values
are taken from the literature.

In order to select candidates that are close enough to the GRB line of sight to produce an
intervening absorption-line-system in the afterglow spectrum, we had to set an upper
limit on the extent of the possible DLA galaxies. Theoretically, it is very difficult to set
a meaningful upper boundary, because different models exist for their galaxy counterparts. 
The observed impact parameter distribution of confirmed DLA galaxies, based on \citet{Peroux2011a}
and \citet{Krogager2012a}, extends from 0.4 to 182 kpc and has a mean value of $\sim25$ kpc.
The impact parameter distribution of DLA galaxy candidates by \citet{Rao2011a} shows similar
characteristics. The majority of DLA galaxies have a small impact parameter, however there are
few cases with large impact parameters ($\sim100\,\rm kpc$).
We therefore follow the statistical approach by \citet{Rao2011a} who used the galaxy number
density as a function of impact parameter as a criterion. They found that the galaxy number
density is comparable to the number density of foreground and background galaxies, i.e. a chance
association is more likely, if the impact parameter exceeds 100 kpc. As a first assumption, we
limit our study to those candidates within 100 kpc from the GRB line of sight, keeping
in mind that this value was derived for intervening DLAs between $z=0.5$ and 0.8.

\subsection{GRB\,050730} \label{sec:grb050730}

GRB\,050730 occurred at a redshift of $z=3.969$. Its afterglow spectrum contains an
intervening sub-DLA ($\log N \left(\rm{\ion{H}{i}}\right)=19.9\pm0.1$,
$\left[\rm{Si}/\rm{H}\right]=-1.5\pm0.2$),\footnote{The metallicity is not corrected for ionisation effects.} and an intervening DLA
($\log N\left(\rm{\ion{H}{i}}\right)=20.3\pm0.1$, $\left[\rm{Si}/\rm{H}\right]<-1.3$) at 
$z=3.02209$ and $z=3.56439$, respectively (Tables \ref{tab:dla_sample}, \ref{tab:metal_sample}). In addition
the afterglow light traversed an intermediate and a strong \ion{Mg}{ii} absorber at $z=2.25313$
and 1.7731, respectively.

In Figure \ref{fig:fov_050730}, we show the field of GRB\,050730 and zoom-ins to the inner
28\farcs8 and 6\farcs0; the region at which the impact parameter exceeds 100 kpc is
highlighted. At the redshift of the aforementioned absorption-line systems, the maximum impact parameter
of 100 kpc translates to an transverse distance between 12\farcs1 and 13\farcs9;
the smaller value belongs to $z=1.7731$ and  the larger to $z=3.56439$, including the astrometric
uncertainty of the afterglow localisation of 0\farcs3.

The GRB was a target of an extensive spectroscopic campaign; several low-resolution spectra
with a total of five different position angles (PA) were obtained with VLT/FORS1 and
VLT/FORS2 (Fig.~\ref{fig:fov_050730}, Table~\ref{tab:log_photometry}). Within the 100-kpc
radius, 19 objects were covered by a slit. In Table~\ref{tab:candidates} we summarise their
magnitudes in different filters and their angular distances from the afterglow position.
Four of these objects were bright enough to allow the extraction of a spectrum. The VLT/FORS1
spectra of objects A1 and A4 are of very low S/N. We detect their continua down to
4500 \AA, but we detect no absorption or emission lines. If these objects
were the galaxy counterparts to the
sub-DLA and DLA, we expect to see the onset of the Ly$\alpha$ forest 
at $\sim4890$ \AA \ and
$\sim5549$ \AA, respectively, which we do not observe. The Ly$\alpha$ non-detection and the extension of
the continuum to even shorter wavelengths, rules them out as the galaxy counterparts.
We detect the continuum of object A7 at very low S/N in the VLT/FORS2 data, but no absorption
or emission lines. The objects A1, A4 and A7 are likely late-type stars based on their colours
and the fact that the morphology and size of their PSFs ($FWHM=0\farcs11$) do not differ from
point sources. Object
A11 lies at the edge of the 100-kpc radius. We detect its continuum down to 4630 \AA \ 
in the VLT/FORS2 data and a drop in flux blueward of it. Assuming that this is the Balmer break,
the redshift of the galaxy is $\sim0.16$, not in agreement with any of the intervening absorbers.

In conclusion, we do not identify any galaxy counterpart candidate of the intervening
sub-DLA and DLA down to a limiting magnitude of $F775W=25.7\,(24.6)\,\rm mag$ and of any of the intermediate
\ion{Mg}{ii} absorbers down to a limiting magnitude of $F775W=26.5\,(25.1) \,\rm mag$, assuming
a maximum impact parameter 50 (100) kpc. These limits were calculated by considering all objects
within 50 (100) kpc but excluding those for which we elucidated the nature or those with a stellar PSF.

\citet{Fynbo2009a} reported the serendipitous discovery of a QSO at the redshift of
$z=3.023$, very similar to the redshift of the sub-DLA in the afterglow spectrum. The QSO
is 17\farcs5 south of the afterglow position (see Fig.~\ref{fig:fov_050730}, Table~\ref{tab:candidates}),
corresponding to a projected distance of $136.7$ kpc at $z=3.02209$, the redshift of the sub-DLA
towards the GRB. It has a brightness
of $20.76\pm0.05$ mag in the $R$-band (corrected for Galactic extinction, Table~\ref{tab:candidates}), but is blended with a
18.2-mag bright K-type star. Strictly speaking, the impact parameter exceeds
the assumed maximum impact parameter of 100 kpc. Given the redshift coincidence 
with the sub-DLA in the afterglow spectrum we discuss this correlated structure
in Sect. \ref{sec:qso_subdla}.

\begin{figure}
\centering
\includegraphics[width=0.5\textwidth, angle=0]{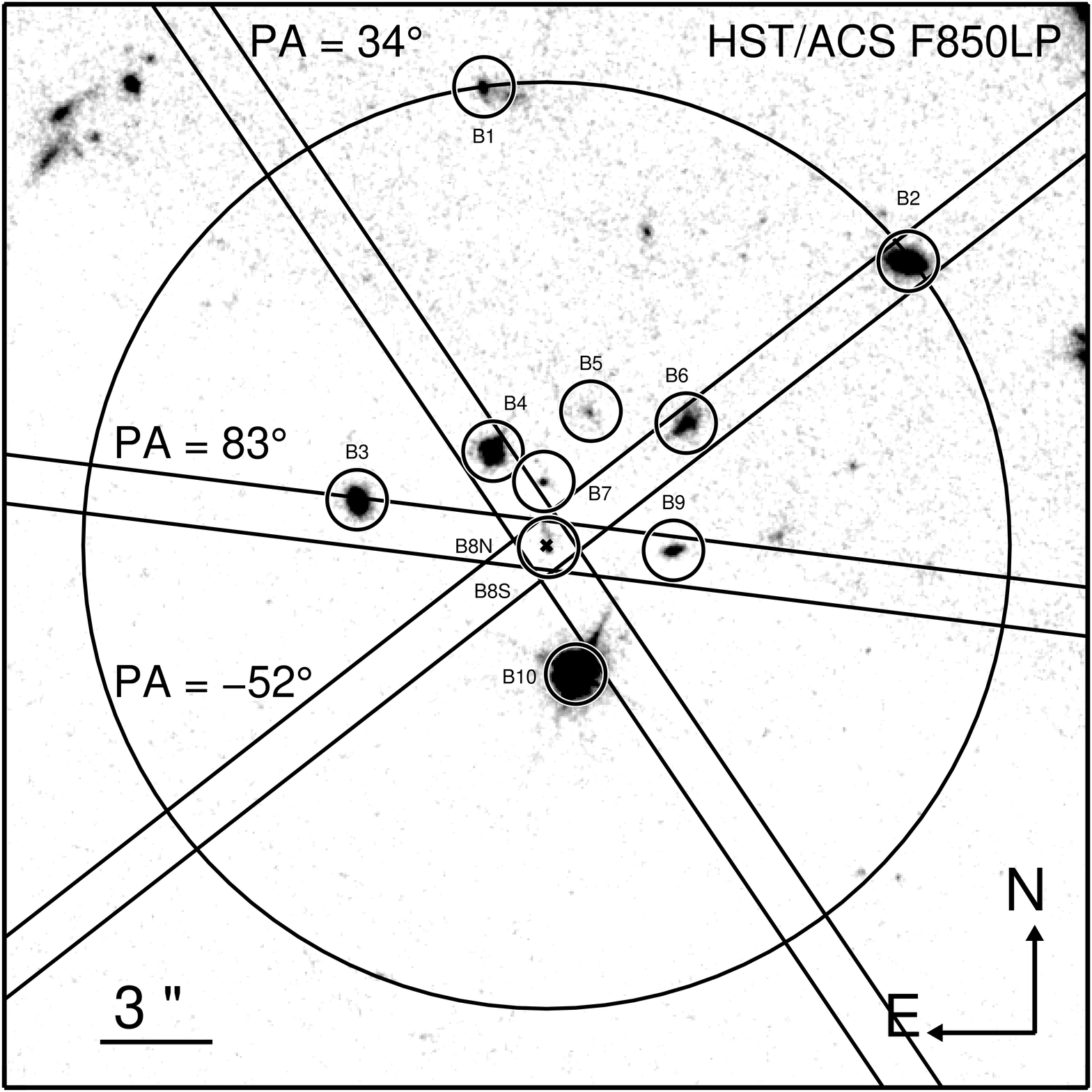}
\caption{Similar to Fig.~\ref{fig:fov_050730}, but for GRB\,050820A ($z=2.615$). In contrast
	to GRB\,050730, the field of GRB\,050820A was observed in 9 filters ranging from 473 nm
	to 5800 nm, in addition to extensive spectroscopy campaigns. This allowed us to construct
	the SED of several objects in the field of view. The width of the displayed field is $29''$,
	twice the maximum impact parameter of 100 kpc of the strong \ion{Mg}{ii} absorber at
	$z=0.6915$, while the circle displays the maximum impact parameter of 100 kpc at the redshift
	of the intervening sub-DLA at $z=2.3598$.
	Every object within 100 kpc is labelled either if it is either detected in at least five filters or
	covered by a slit. Objects B7 and B10 appear to be not in the slit, due to the smaller
	PSF of the \hst \ in comparison to the VLT.
	}
\label{fig:fov_050820A}
\end{figure}

\subsection{GRB\,050820A}\label{sec:050820A}

\citet{Ledoux2005a} and \citet{Vergani2009a} reported an intervening sub-DLA at $z=2.3598$ with
$\log N\left(\rm{\ion{H}{i}}\right)=20.1\pm0.2$ and a metallicity of $\left[\rm{Si}/\rm{H}\right]=-1.5\pm0.2$
and two strong \ion{Mg}{ii} absorbers between $z=0.6915$ and 1.6204
towards GRB\,050820A ($z_{\rm GRB} = 2.615$, Tables~\ref{tab:dla_sample}, \ref{tab:metal_sample})

In Figure \ref{fig:fov_050820A}, we show the inner $29''$ around the afterglow position
and highlight the region at which the impact parameter is 100 kpc. At the redshift
of the intervening absorbers, the maximum impact parameters correspond to 12\farcs4 for $z=2.3598$
and 14\farcs4 for $z=0.6915$, including the uncertainty of 0\farcs4 in the afterglow localisation.

\begin{figure}
\centering
\includegraphics[bb=27 40 545 318, clip, width=0.5\textwidth, angle=0]{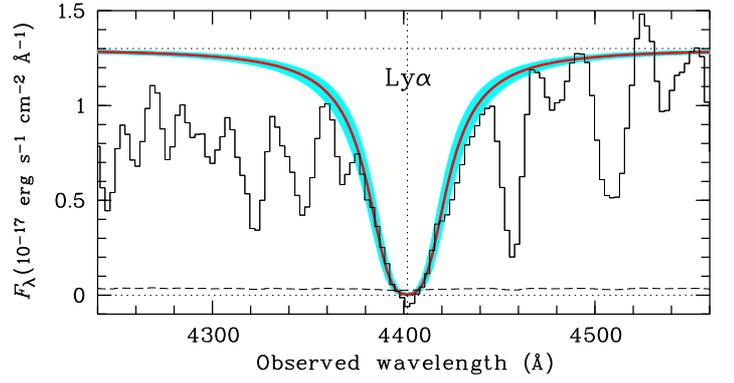}
\caption{ 
The GRB\,050908 afterglow spectrum, obtained with VLT/FORS1, centred on the intervening
Ly$\alpha$ absorption line ($z_{\rm DLA}=2.6208$; $\log N(\ion{H}{i})=20.8\pm0.1$).
A neutral hydrogen column density fit to the damped Ly$\alpha$  line is shown with a solid
line, while the shaded region indicates the $1\sigma$ errors.
}
\label{fig:lyalpha_050908}
\end{figure}

\begin{figure}
\centering
\includegraphics[width=0.5\textwidth, angle=0]{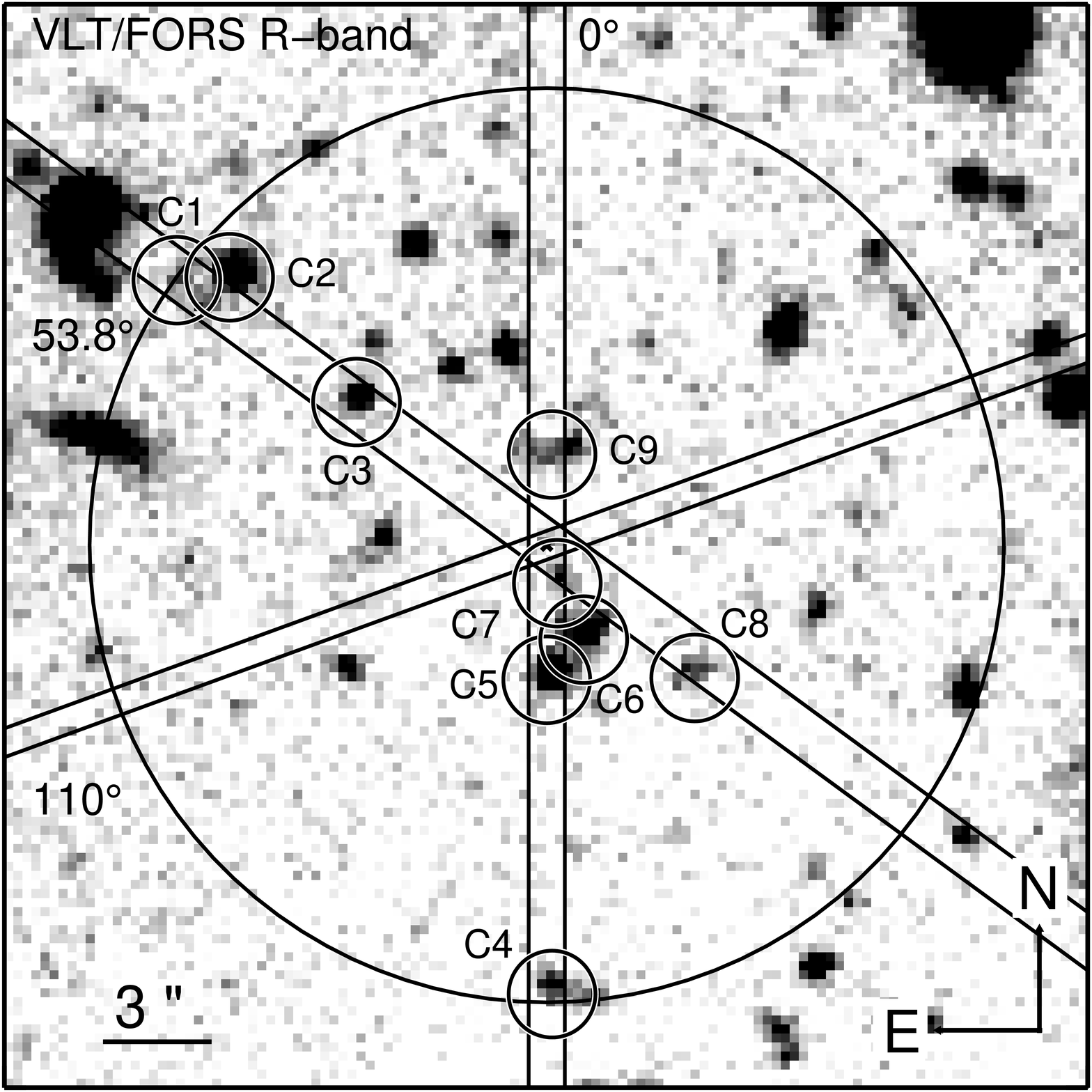}
\caption{Similar to Fig.~\ref{fig:fov_050730}, but for GRB\,050908 ($z=3.3467$). 
	The different slit orientations from Table~\ref{tab:log_photometry}
	are over-plotted. The circle radius of 12\farcs7 represents the assumed
	maximum impact parameter of 100 kpc of the intervening sub-DLA at $z=2.6208$.
	Objects are labelled if they lie within the circle and if they
	are covered by any of the slits. Their magnitudes are summarised
	in Table~\ref{tab:candidates}.}
\label{fig:fov_050908}
\end{figure}

After the afterglow faded, several spectra were acquired with VLT/FORS2. In
addition, the field was also the target of an extensive photometric campaign
covering nine filters from 473.1 nm ($g'$-band) to $5.8\,\mu\rm m$;
for full details we refer to Sect.~\ref{sec:selection} and Table
\ref{tab:log_photometry}. This allows us to elucidate the nature of several
objects within 100 kpc, not only of those that were covered by a slit and
bright enough for a spectrum to be extracted. In the following, we only consider
those objects that are either detected in at least five filters to attempt
SED modelling, or fall in one of the slit positions. In total, 11
objects within 100 kpc of the afterglow position fulfil these criteria (Fig.
\ref{fig:fov_050820A}, Table~\ref{tab:candidates}). Among them, 9 fell into 
one of the slits. 

The two brightest objects (B4 and B10) within the 100 kpc radius are late-type
stars.  The spectrum of B9 exhibits two emission lines at 5322.1 and
7149.3 \AA \ both detected at $\sim6\sigma$ confidence level 
(Fig.~\ref{fig:GRB080520A_B09}).
We identify these lines as [\ion{O}{ii}] $\lambda$3727 and [\ion{O}{iii}]
$\lambda$5007 at a common redshift of $z=0.428$.

The spectrum of object B2 (Fig.~\ref{fig:GRB050820A_B02})
has a continuum break at $\sim6750$ \AA~and absorption lines at
6660.2 and 6719.5 \AA. These features are consistent with the Balmer break
and \ion{Ca}{ii} K\&H absorption at a redshift of $z=0.693$. 
The S/N ratio of the VLT/FORS2 spectra of B3 and B6 are too low for redshift
determination.\footnote{The spectrum of B3 shows a prominent emission line at
$\sim6310\,\AA$. The bright [\ion{O}{i}] sky emission line at 6300.3\,\AA~partly
overlaps with this feature, making the identification ambiguous. 
If this is indeed an emission line of galaxy B3, it is likely
$\left[\ion{O}{ii}\right]\,\lambda3727$ redshifted to $z\sim0.693$,
coinciding in redshift with a strong \ion{Mg}{ii} absorber.}
Both objects are detected in the nine filters, see
Fig~\ref{fig:GRB050820A_B03} and 
Fig.~\ref{fig:GRB050820A_B06}. The best fits to their SEDs suggest
that B3 and B6 are galaxies at
$z=0.67^{+0.05} _{-0.06}$ and $1.46^{+0.07} _{-0.41}$, respectively.  Object B5 was detected in
five filters from 473 to 766 nm (Fig. \ref{fig:GRB050820A_B05}). The SED
is best described by a young and small galaxy at $z=0.9^{+0.1} _{-0.2}$. We caution
that the solution is not unique because the galaxy is only detected in five filters.
The redshifts of B2, B3, B5 and B6 match the redshifts of the strong \ion{Mg}{ii}
absorbers at $z=0.6915$ and 1.4288 towards GRB\,050820A, see Table \ref{tab:metal_sample}.

The continua of objects B7 and B8 are visible in the 2D spectrum. Due to seeing
the spectrum of B8 is not resolved into the two objects B8N
and B8S as seen in the \hst~image. The extracted 1D spectra of B7 and B8 have a
low S/N ratio and thus their redshifts cannot be determined. Recently,
\citet{Chen2011a} observed the field with the IR echellete spectrograph FIRE
mounted on the Magellan telescope. The slit was oriented to cover object B7
and the compound B8. Based on these observations, \citet{Chen2011a} found
that the galaxy B7 and the host complex B8 are a group of galaxies at $z\simeq2.613$,
in contrast to \cite{Chen2009a} who suggested that B7 and B8S are interacting
galaxies forming the strong \ion{Mg}{ii} absorber at $z=0.692$.

Although we elucidated the nature of all objects within 3\farcs7
that are visible in the \hst~images, we did not identify a possible galaxy counterpart of
the intervening sub-DLA. We also did not find any candidate brighter than
$F625W=26.7\,(26.6)\,\rm mag$ with an maximum impact parameter of 50 (100) kpc.

\subsection{GRB\,050908}

\begin{figure}
\centering
\includegraphics[bb=39 40 545 767, clip, width=0.5\textwidth, angle=0]{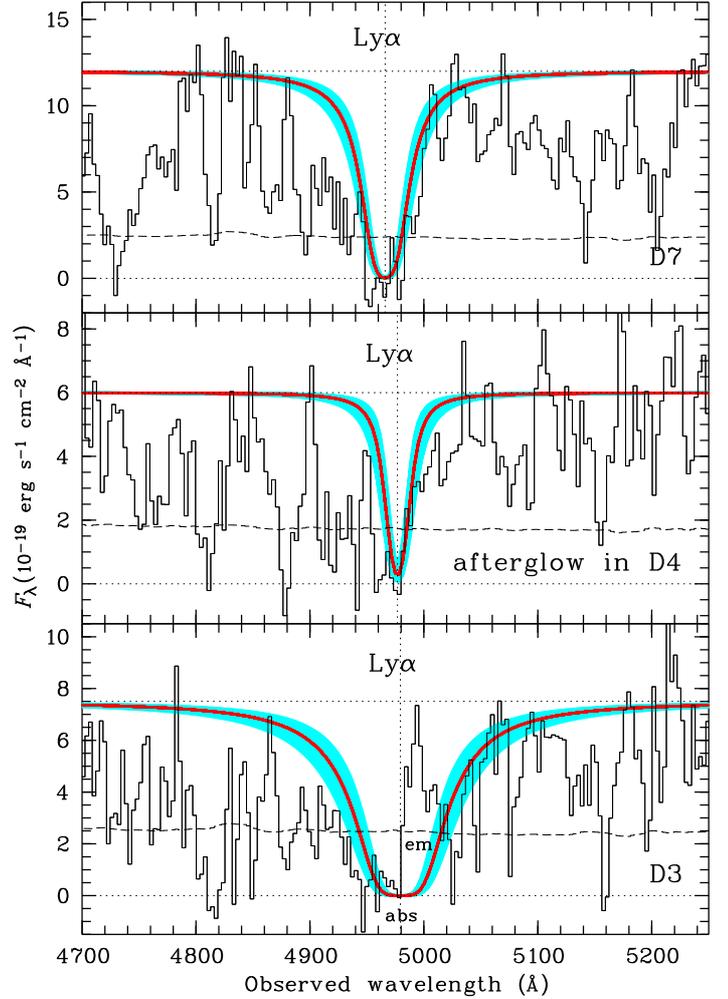}
\caption{Fit of the intervening Ly$\alpha$ absorption features towards galaxy D7 ($z_{\rm abs} = 3.085$; $\log N(\ion{H}{i})=20.7\pm0.2$)
   and towards GRB\,070721B ($z_{\rm abs} = 3.0939$; $\log N(\ion{H}{i})=20.1\pm0.3$), and
   the Ly$\alpha$ absorption feature in the DLA galaxy D3 ($z_{\rm abs} = 3.096$; $\log N(\ion{H}{i})=21.3\pm0.2$). The fit of D3
   nicely shows the Ly$\alpha$ emission in the red part of the trough. The afterglow spectrum (D4)
   was acquired with VLT/FORS2, while the displayed spectra of D3 and D7 were extracted from the VLT/FORS1 data.
   The fit is shown with a solid line, while 1$\sigma$ errors are displayed with the shaded region.
}
\label{fig:grb070721B_dla_1}
\end{figure}

\begin{figure}
\centering
\includegraphics[width=0.5\textwidth, angle=0]{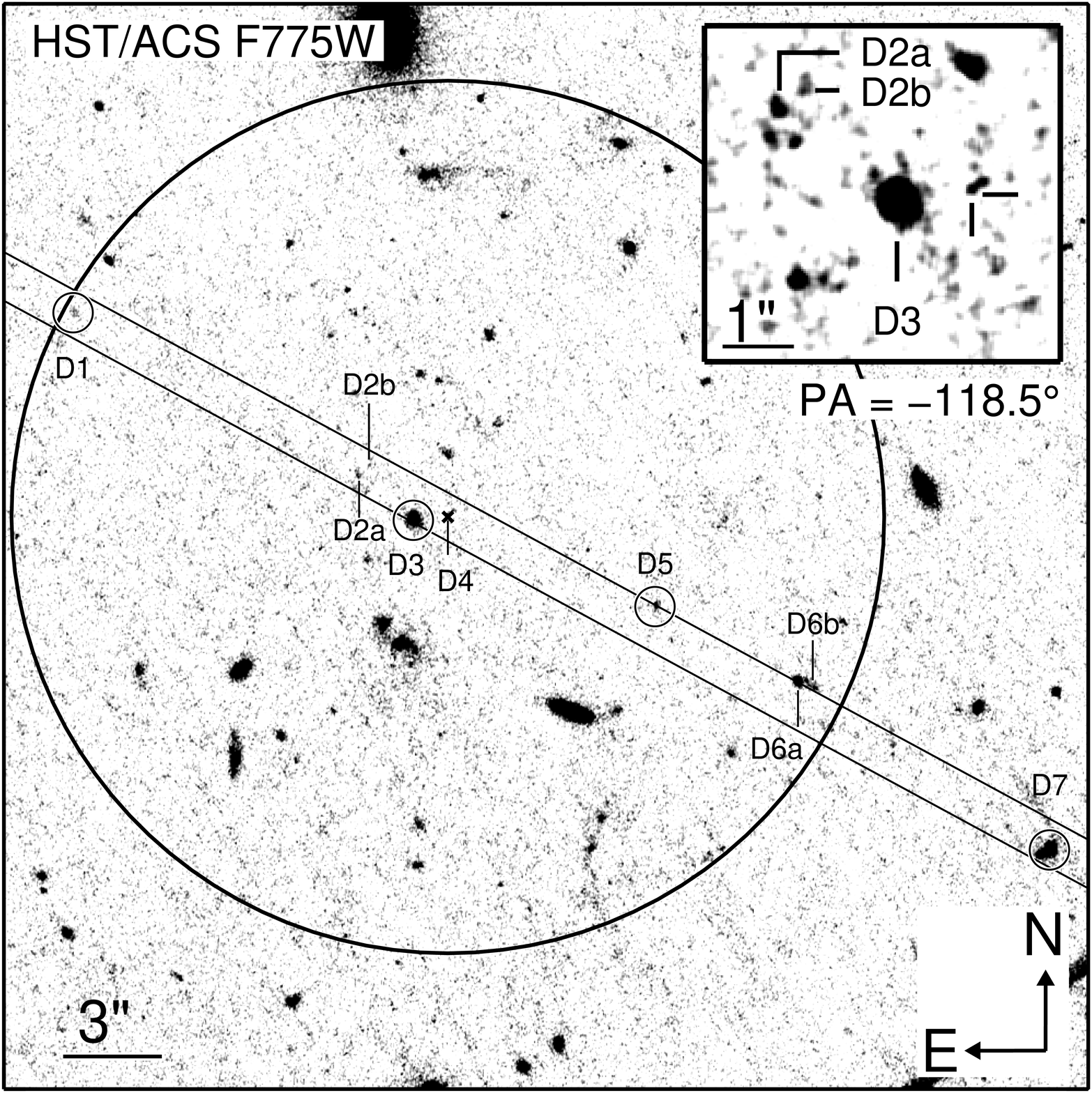}
\caption{Similar to Fig.~\ref{fig:fov_050730}, but for GRB\,070721B ($z=3.6298$).
	Object D4 is the host galaxy. Object D3 is the galaxy counterpart to the
	intervening DLA ($z=3.094$) towards GRB\,070721B. The impact parameter
	of the DLA galaxy is 1\arcsec (7.9 kpc). The line of sight
	of object D7 also traverses a DLA at $z=3.085$. The distance between
	objects D3 and D7 is 21\farcs6, corresponding to a projected distance of
	167.4 kpc at $z=3.09$. The inset is a 5\arcsec \ zoom-in on the position of D3.
	The afterglow position is marked by the cross-hair. 
	}
\label{fig:fov_070721B}
\end{figure}

\begin{figure*}[th!]
\centering
\includegraphics[bb= 10 2 1245 452, clip, width=1.0\textwidth, angle=0]{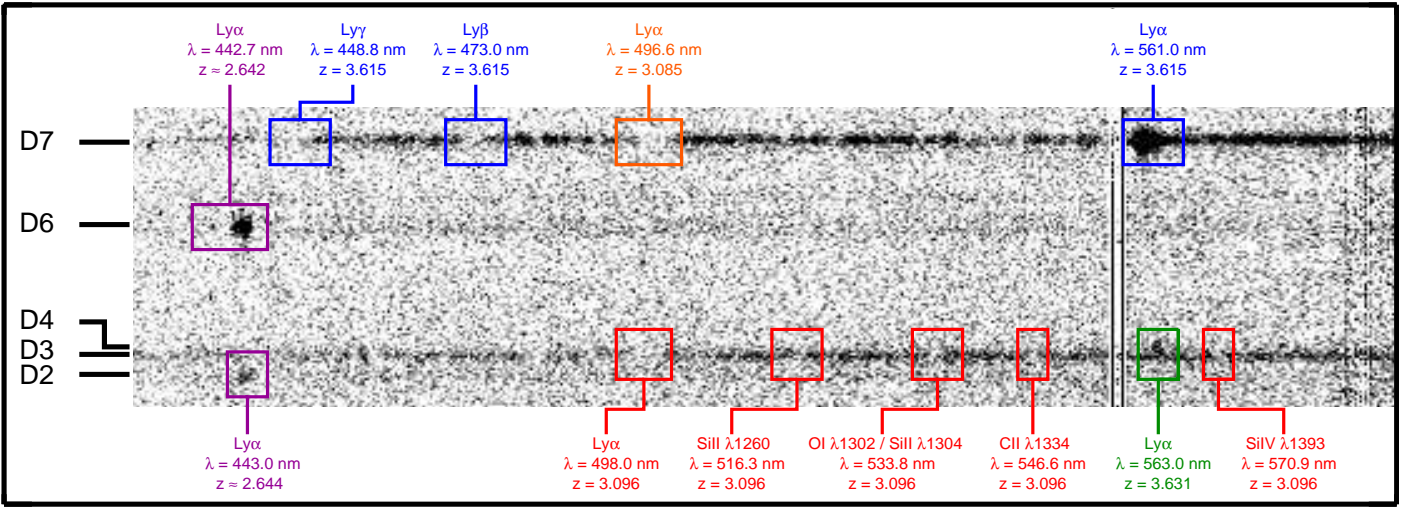}
\caption{The VLT/FORS1 2D spectrum obtained after the optical afterglow of GRB\,070721B faded. The photometric counterparts
	of the different traces are labelled according to Fig.~\ref{fig:fov_070721B} and Table~\ref{tab:candidates}.
	The host (D4) is only visible via  its Ly$\alpha$ emission. The afterglow spectrum of GRB\,070721B
	would be located at the position of D4. Prominent absorption and emission lines of the galaxies
	and intervening absorption-line systems are highlighted and color coded.
	The intervening DLA present in the afterglow spectrum is the very bright galaxy D3.
	The spectrum of the Ly$\alpha$ emitter D7 displays an intervening DLA as well.
	}
\label{fig:070721B_2d_spectrum}
\end{figure*}

The afterglow spectrum of GRB\,050908 ($z=3.3467$), obtained with VLT/FORS1,
revealed an intervening DLA at $z=2.6208$ with $\log N\left(\rm{\ion{H}{i}}\right)=20.8\pm0.1$,
as derived from the Voigt profile fit shown in Fig.~\ref{fig:lyalpha_050908}.
We thus revise the conclusion by \citet{Fynbo2009a} that this intervening absorber
is a sub-DLA. This DLA is peculiar for its strong metal lines ($EW_{\rm
rest}\left(\rm{\ion{Si}{ii}}\,\lambda1526\right)=2.24\pm0.06\,\rm \AA$) that are
stronger than those of most intervening DLAs \citep[][their Fig.~11]{Fynbo2009a}.
In addition, the afterglow spectrum comprises of an intermediate \ion{Mg}{ii}
absorber at $z=1.5481$. The properties of both absorbers are summarised in Tables
\ref{tab:dla_sample} and \ref{tab:metal_sample}.

At the redshift of the intervening \ion{Mg}{ii} absorber and the DLA, an
impact parameter of 100 kpc corresponds to an angular distance of 12\farcs0
and 12\farcs6, respectively, including the uncertainty in the afterglow position
of 0\farcs3. 
The GRB field and the galaxies within this impact parameter
are shown in Fig.~\ref{fig:fov_050908}.

An afterglow spectrum was acquired with VLT/FORS1 ($\rm{PA}=0\fdg0$) and with
Gemini-N/GMOS ($\rm{PA}=110\fdg0$). In addition, a further spectrum was
obtained with VLT/FORS1 ($\rm{PA}=53\fdg8$) two years later
(Table~\ref{tab:log_photometry}). Both FORS1 spectra cover several faint objects
($R\sim26\,\rm mag$) within the 100-kpc radius (Fig.~\ref{fig:fov_050908},
Table~\ref{tab:candidates}). Apart from the afterglow, we only detect a very weak
emission feature at $\sim4512$ \AA \ at the position of C1. Assuming that this
is a genuine emission line, the line is likely Ly$\alpha$ redshifted to $z\sim2.71$,
and not $\left[\rm{\ion{O}{ii}}\right]$ because of the non-detection of H$\beta$
and  $\left[\rm{\ion{O}{iii}}\right]\,\lambda5007$.
The Gemini spectrum ($\rm{PA}=110\fdg0$) does not cover any of the faint objects
visible in the $R$-band image and we do not detect any emission line in these data.

Since the field is quite crowded and the amount of spectroscopic and multi-filter
data are limited, we can only place a shallow upper limit on the brightness of the
galaxy counterpart of the both absorption-line systems. Assuming a maximum impact
parameter of 50 or 100 kpc, the galaxy counterpart cannot be brighter than
$25\,\rm mag$ and 24.7 mag in the $R$ band.

\subsection{GRB\,070721B}\label{sec:res_070712B}

The VLT/FORS2 afterglow spectrum of GRB\,070721B ($z=3.6298$) revealed an
intervening absorber at $z=3.0939$, a sub-DLA with strong metal absorption
lines ($EW_{\rm rest}\left(\rm{\ion{Si}{ii}}\,\lambda1526\right)=1.71\pm0.54\,\rm \AA$;
Table~\ref{tab:dla_sample}). Figure \ref{fig:grb070721B_dla_1} displays the
Ly$\alpha$ absorption profile; the Voigt profile fit gives $\log\,N\left(\rm{\ion{H}{i}}\right)=20.1\pm0.3$.
In Figure \ref{fig:fov_070721B} we show the field of view and highlight the 
maximum impact parameter of 100 kpc, corresponding to 13\farcs3 at $z=3.09$,
including the uncertainty in the afterglow localisation of 0\farcs4.

Four months after the GRB, an additional spectrum with $\rm{PA}=-118\fdg5$ was
acquired with VLT/FORS1. The spectrum covers several objects within the 100 kpc radius
(Table~\ref{tab:candidates}). The 2D spectrum shown in
Fig.~\ref{fig:070721B_2d_spectrum} reveals that this is a very complex and
perplexing line-of-sight.  The galaxy D7, which was also detected in the
2D-spectrum of the afterglow, is a bright Ly$\alpha$ emitter at $z=3.615$ -
very similar to the redshift of the GRB. Object D4 is the host galaxy of
GRB\,070721B and is only detected by its Ly$\alpha$ emission line in the 2D
spectrum (Fig.~\ref{fig:070721B_2d_spectrum}). The emission line redshift
($z_{\rm em}=3.631$) differs slightly from the absorption-line redshift
($z_{\rm abs}=3.6298$), derived from metal lines in the afterglow spectrum (see
\citealt{MilvangJensen2012a} for a detailed discussion). In the spectrum
of the galaxy D3 we detect several strong metal lines as well as a strong
Ly$\alpha$ absorption line at a common redshift of $z=3.096$. This galaxy has
an impact parameter of only 1\arcsec relative to the GRB position. Hence,
D3 must be the galaxy counterpart of the DLA seen in the GRB afterglow
spectrum. 
This was also suggested by \citet{Fynbo2009a} and \citet{Chen2009a}.
Intriguingly, D7 also has an intervening strong Ly$\alpha$ absorption
line at a very similar redshift ($z=3.085$); we derive a \ion{H}{i}
column density of $20.7\pm0.2$ from the Voigt profile fitting, as
shown in Fig. \ref{fig:grb070721B_dla_1}.\footnote{There are some uncertainties in
the Ly$\alpha$ profile fits for D3, D4 and D7. Because of the resolution and
the quality of the spectra, the damped wings are barely visible (see Fig.~\ref{fig:grb070721B_dla_1}). 
We cannot rule out the possibility
that a blend of narrow Ly$\alpha$ lines mimics the strong absorption feature.
It is not very likely, because it would require that D3 and D7 have a
comparable density of \ion{H}{i} clumps at almost identical redshifts.
Strictly speaking, the stated column densities are only upper limits. The fit
does not rule out an extremely large Doppler parameter
($b\gg100\,\rm{km}\,\rm{s}^{-1}$), implying a reduction of the \ion{H}{i}
content by a factor of 100. This remark has in particular to be kept in mind
for the strong Ly$\alpha$ absorption feature of the intervening absorber towards
D7.} However, we do not detect metal lines associated with this intervening absorption
line system.\footnote{We obtain the following $3\sigma$ upper limits from the
VLT/FORS1 spectrum, assuming an aperture of $1665\,\rm{km\,s}^{-1}$ (twice the $FWHM$ of
a Gaussian with a Doppler parameter of $b=500\,\rm{km\,s}^{-1}$):
$EW_{\rm{rest}}\left(\rm{\ion{Si}{ii}}\,\lambda1260\right)<3.6\,\rm \AA$,
$EW_{\rm{rest}}\left(\rm{\ion{O}{i}}\,\lambda1302\right)<2.8\,\rm \AA$,
$EW_{\rm{rest}}\left(\rm{\ion{C}{ii}}\,\lambda1334\right)<2.8\,\rm \AA$,
$EW_{\rm{rest}}\left(\rm{\ion{Si}{iv}}\,\lambda1339\right)<2.7\,\rm \AA$,
$EW_{\rm{rest}}\left(\rm{\ion{C}{iv}}\,\lambda1548\right)<2.7\,\rm \AA$,
$EW_{\rm{rest}}\left(\rm{\ion{Fe}{ii}}\,\lambda1608\right)<3.0\,\rm \AA$, and
$EW_{\rm{rest}}\left(\rm{\ion{Al}{ii}}\,\lambda1670\right)<3.3\,\rm \AA$. These limits,
are on average less stringent than the 
measurements of the detected metal lines in the DLA galaxy D3 (see Table \ref{tab:dla_linelist}).}
At that redshift the angular separation between D3 and D7 of 20\farcs7 translates
into a transverse distance of 161 kpc.
The weak trace below D7 is from the galaxy D6. This spectrum displays a single
strong emission line. It is very likely Ly$\alpha$ at $z_{\rm em}=2.642$ and
not $\left[\ion{O}{ii}\,\lambda3727\right]$, because of the lack of H$\beta$ and
$\left[\rm{\ion{O}{iii}}\right]\,\lambda5007$ in emission. The
spectrum of D2 shows that this is also a Ly$\alpha$ emitter at a very similar redshift
to that of D6, $z_{\rm
em}\simeq2.644$. In the spectrum of the afterglow we detect absorption
lines from Ly$\alpha$, 
$\rm\ion{C}{ii}\,\lambda1334$ and the
\ion{C}{iv} doublet (unresolved) at $z_{\rm abs}=2.655$ albeit a low S/N ratio.
The velocity distance between the absorption and emission line
redshifts is $\sim900\,\rm{km}\,\rm{s}^{-1}$. 

The region around Ly$\alpha$ in the spectrum of D3 seems to be a superposition
of an emission line on top of a broad absorption line (see
Figs.~\ref{fig:grb070721B_dla_1}, \ref{fig:070721B_2d_spectrum}).  Such
a feature has been observed in several high-$z$ galaxies
\citep[e.g.][]{Pettini1998a, Pettini1998b, Pettini2000a}.  We estimate an
\ion{H}{i} column density of $\log\,N\left(\ion{H}{i}\right)=21.3\pm0.2$.

\section{Discussion}\label{sec:discussion}
\subsection{Detected galaxy counterparts}
\subsubsection{DLA galaxy towards GRB\,070721B}

In Sect.~\ref{sec:res_070712B} we showed that the galaxy counterpart of the intervening DLA
(object D3, $z_{\rm D3}=3.096\pm0.003$) has an impact parameter of 7.9 kpc and an
extinction-corrected $R$-band magnitude of $24.41\pm0.02$ mag (Fig.~\ref{fig:fov_070721B},
Table~\ref{tab:candidates}). The co-ordinates are $RA\rm{(J2000)}=02^{\rm h}12^{\rm m}33\fs018$ and
$DEC\rm{(J2000)}=-02^\circ11'40\farcs99$ with an uncertainty of 0\farcs4 in each coordinate. We denote the object
as DLA J0212-0211 in the following.
Up to now, only one further sub-DLA/DLA galaxy was identified beyond
$z>3$. \citet{Djorgovski1996a} found the galaxy counterpart ($R=24.8\pm0.2\,\rm mag$)
to the $z=3.15$ sub-DLA ($\log N(\ion{H}{i})=20.0$; \citealt{Lu1993a}) towards QSO B2233+131.
Compared to typical $z\simeq3$ galaxies, both objects are comparable to the brightness of
an $L_*$ galaxy ($R_{*} = 24.48\pm0.15\,\rm mag$; \citealt{Steidel1999a}).

At low-$z$ ($z<1$), the overwhelming majority (75\%) of sub-DLA/DLA galaxies summarised
in \citet{Peroux2011a} are fainter than both $z>3$ sub-DLA/DLA galaxies. The mean luminosity of
that sample is $0.66\,L_*$ and the median is $0.32\,L_*$. The sample is not tightly distributed
around the mean and the brightest galaxy in that ensemble reaches $2.8\,L_*$. A proper comparison
with high-z ($z>2$) DLA galaxies is not possible. To date, the luminosity of only 4 of 10
sub-DLA/DLA galaxies is known. The reason for this is that most of them were detected by their
Ly$\alpha$ emission and have small impact parameters so that they are outshone by the glare of
the quasar. This hampers the determination of their luminosity.
\citet{Chen2009a} concludes that the DLA galaxy in the field of GRB\,070721B is the
most luminous DLA galaxy at $z>2$. The morphology of the galaxy is undisturbed, based on the
shape in the \hst~image, inset in Fig. \ref{fig:fov_070721B}. We measure an ellipticity of 0.15
and a half-light radius of 1.2 kpc, obtained with \texttt{SExtractor}. The non-detection
of a galaxy interaction or merger is not surprising. \citet{Overzier2010a} argued that a detection
is hampered at $z\sim3$ because of the reduced physical resolution and sensitivity in addition
to the general difficulties of observing interacting galaxies.

\begin{figure}[t!]
\centering
\includegraphics[bb= 2 1 687 629, clip, width=0.49\textwidth, angle=0]{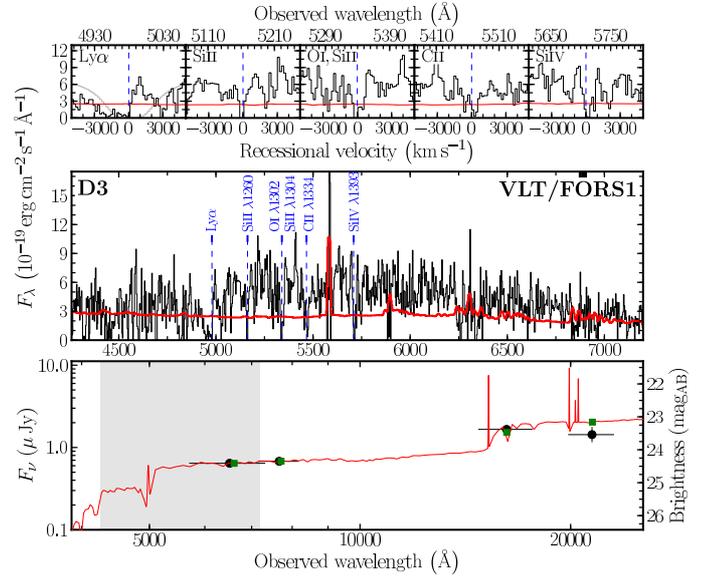}
\caption{Spectrum (middle) and the SED (bottom) of the DLA galaxy D3 in the field of GRB\,070721B.
	In the middle panel, the error spectrum is overplotted and several absorption lines are marked. 
	Regions of strong telluric features (atmosphere transparency $< 20\%$) were not used in
	the spectral analysis and are marked by a small box at the top. The top panel shows zoom-ins
	on the individual absorption lines.
	In the bottom panel, the observed extinction-corrected data points are shown as circles with
	error bars. The curve represents the best fit to the observed SED. The model predicted magnitudes
	(squares) are superposed. The gray area highlights the interval that is covered by the spectrum above.
	}
 \label{fig:GRB070721B_D03}
\end{figure}

The Ly$\alpha$ absorption line shows an excess of flux in the red part of the absorption
profile (Fig.~\ref{fig:grb070721B_dla_1}; Sect.~\ref{sec:res_070712B}). The extracted
emission line profile, shown in Fig. \ref{fig:GRB070721B_D03}, is slightly
asymmetric and peaks at $\sim1000\,\rm{km}\,\rm{s}^{-1}$ with respect to the systemic redshift.
We measure a line flux density of
$\left(2.34 \pm0.25\right)\times 10^{-17}\,\rm{erg}\,\rm{cm}^{-2}\,\rm{s}^{-1}\,\rm{\AA}^{-1}$,
i.e. a line significance of 9.2 $\sigma$. 
\citet{Pettini1998a, Pettini1998b, Pettini2000a} reported the detection of the same
feature with peak recessional velocities between $+400$ and $1100\,\rm{km}\,\rm{s}^{-1}$
in several $z\approx3$ galaxies \citep[see also][]{Adelberger2003a}. They argue that this feature is similar
to a P-Cygni profile, indicating the presence of a galactic outflow (also found in starburst
galaxies; e.g. \citealt{Kunth1998a, GonzalezDelgado1998a}; but see also \citealt{Verhamme2006a,Laursen2009a}).

To further explore the properties of the galaxy counterpart, we fit the SED with \texttt{Le PHARE}
(Sect. \ref{sec:sed_fitting}), as displayed in Fig.~\ref{fig:GRB070721B_D03}.
Leaving all model parameters free, except for the redshift, which was fixed to $z=3.096$, the SED is
best described by a young ($0.4\,\rm Gyr$) and close to dust-free galaxy
($A_V =0.3\,\rm mag$, Calzetti reddening; see Table \ref{tab:grb050820A_mgii} for all best fit values).
The SFR of $37\,M_{\sun}\,\rm{yr}^{-1}$ is similar to the that of the $z=3.15$
DLA galaxy \citep{Djorgovski1996a,Christensen2004a, Peroux2011a} and similar to the SFR of
typical $U$-band dropout LBGs \citep{Giavalisco2002a}.

\begin{table}[t!]
\caption{Properties of the galaxy counterpart candidates of the strong \ion{Mg}{ii}
absorber towards GRB\,050820A ($z_{\rm{\ion{Mg}{ii}}}=0.692$ and 1.430) and the DLA
galaxy towards GRB\,070721B ($z_{\rm DLA}=3.094$)}
\scriptsize
\centering
\begin{tabular}{l@{\hspace{3.5mm}}c@{\hspace{3.5mm}}c@{\hspace{3.5mm}}c@{\hspace{3.5mm}}c@{\hspace{3.5mm}}c}
% \begin{tabular}{lccccc}
\toprule
						& \multicolumn{4}{c}{GRB\,050820A}							& GRB\,070721B\\
						& B2				& B3				& B5			& B6				& D3\\
\midrule
\vspace{0.5mm}$z$				& 0.693				& $0.67^{+0.05} _{-0.06}$	& $0.9^{+0.1} _{-0.2}$	& $1.46^{+0.07} _{-0.41}$	& 3.096\\
\vspace{0.5mm}$\theta\,(\rm{kpc})$		& 87.5				& 36.8				& 26.4			& 42.0				& 7.9\\
\vspace{0.5mm}$M_B\,(\rm{mag})$			& $-20.2$			& $-19.4$			& $-17.7$		& $-21.4$			& $-22.3$\\
\vspace{0.5mm}$L/L_*$				& 0.3				& 0.2				& 0.03			& 0.9				& $\sim1$\\
\vspace{0.5mm}Dust model			& \multicolumn{5}{c}{Calzetti}\\	
\vspace{0.5mm}$A_{\rm V}\,\left(\rm{mag}\right)$& 0.3				& 1.6				& 0.3			& 0.3				& $0.3$\\
\vspace{0.5mm}Age (Gyr)				& $2.0\pm0.2$			& $1.3^{+2.3} _{-0.7}$		& $0.6^{+1.3} _{-0.4}$	& $0.5^{+0.5} _{-0.1}$		& $0.4^{+0.5} _{-0.2}$\\
\vspace{0.5mm}$\log\,M/M_{\sun}	$		& $10.3\pm0.1$			& $9.9^{+0.2} _{-0.1}$		& $8.4\pm0.2$		& $10.1\pm0.1$			& $10.1^{+0.3} _{-0.2}$\\
\vspace{0.5mm}$SFR \,\left(M_{\odot}\,\rm{yr}^{-1}\right)$& $0.15^{+0.03} _{-0.02}$& $7.5\pm2.7$		& $0.5^{+1.2} _{-0.3}$	& $1.5^{+1.5} _{-0.6}$		& $37^{+117} _{-23}$\\
$\chi^2/\rm{n.o.f.}$\tablefootmark{a}		& 16.3/9			& 7.3/9				& 1.5/5			& 1.6/9				& 5.0/4\\
\bottomrule
\end{tabular}
\tablefoot{The table summaries the most important properties
	of the SED fits of the most likely candidates. The redshifts of B2 and D3 were fixed to the spectroscopically
	measured redshift in the SED fit. The impact parameters, $\theta$, are	calculated from Table \ref{tab:candidates},
	assuming either $z=0.6915$, 1.430, or 3.096. The luminosities were computed using the results by \citet{Dahlen2005a}
	and \citet{Steidel1999a} on LFs; the knee of the luminosity functions $L_*$ is at $M_{\rm{B},\,*}=-21.43$
	$M_B=-21.60$ and $m_R=24.48$ mag at $z=0.692$, 1.4288 and 3.096, respectively. We note that the stellar mass and
	its uncertainty was derived assuming a Chabrier IMF \citep[][see also Sect. \ref{sec:sed_fitting} for more details on
	the SED fit]{Chabrier2000a}.
	\tablefoottext{a} "n.o.f." stands for number of filters.}
\label{tab:grb050820A_mgii}
\end{table}

\begin{table*}
\caption{DLA J0212-0211 absorption lines in the DLA galaxy and the GRB 070721B afterglow spectrum}
\centering
\begin{tabular}{cc@{\hspace{1cm}}cccc@{\hspace{0.3cm}}ccc}
\toprule
\multirow{2}{*}{Ion}	& Transition	&$EW_{\rm rest}$\tablefootmark{a}	& \multirow{2}{*}{$\log N$}	&  \multirow{2}{*}{$\left[\rm{X}/\rm{H}\right]$}	&&$EW_{\rm rest}$\tablefootmark{a}&\multirow{2}{*}{$\log N$}	&  \multirow{2}{*}{$\left[\rm{X}/\rm{H}\right]$}\\
			& (\AA)		&(\AA)					& 				&			&& (\AA)		& 				&		\\
\midrule
			&		&\multicolumn{3}{c}{\textbf{Spectrum of the DLA galaxy D3}}					&& \multicolumn{3}{c}{\textbf{D3 in the afterglow spectrum}}\\
\midrule
\ion{H}{i}     	 	&    1215       &                                   	& $21.3\pm0.2$\tablefootmark{b} &               	&&                          		& $20.1\pm0.3$\tablefootmark{b}	&  \\
\ion{Si}{ii}   		&    1260       &    $3.00\pm0.98$                  	& $>14.32$\tablefootmark{c}     &    $>-2.49$           &&    $<2.44$               		& \dots             		& \dots   \\
\ion{O}{i} / \ion{Si}{ii}& 1302/1304   	&    $3.94\pm1.15$                  	& \dots                      	&    \dots              &&    $3.06\pm0.83$         		& \dots              		& \dots   \\
\ion{C}{ii}     	&    1334       &    $3.97\pm1.11$                  	& $>15.29$			&    $>-2.44$           &&    $<3.57$              		& \dots            		& \dots   \\
\ion{Si}{iv}    	&    1393       &    $3.41\pm1.01$                  	& $>14.57$          		&    \dots              &&    $<2.59$               		& \dots             		& \dots	  \\
\ion{Si}{ii}  		&    1526       &    $<5.05$                        	& \dots                      	&    \dots              &&    $2.35\pm0.69$         		& $>14.95$\tablefootmark{c}	& $>-0.66$\\
\ion{C}{iv}    		&    1548/1549  &    $<6.13$                        	& \dots                      	&    \dots              &&    $4.34\pm0.73$         		& $>15.03$          		& \dots   \\
\ion{Fe}{ii}   		&    1608       &    $<4.98$                        	& \dots                      	&    \dots              &&    $3.02\pm0.62$         		& $>15.39$          		& $>-0.21$\\
\ion{Al}{ii}   		&    1670       &    $<7.33$                        	& \dots                      	&    \dots              &&    $2.33\pm0.74$         		& $>13.70$           		& $>-0.85$\\
\bottomrule
\end{tabular}
\tablefoot{The last column denotes the metal abundance with respect to solar.
\tablefoottext{a} The restframe $EW$ ($1\sigma$ errors) and the $3\sigma$ detection limits are measured assuming
an aperture of $1665\,\rm{km}\,\rm{s}^{-1}$, i.e. twice the full width at half maximum of a Gaussian with Doppler parameter
$b=500\,\rm{km\,s}^{-1}$.
\tablefoottext{b} Derived from the Voigt profile fit of the Ly$\alpha$ absorption, see Fig. \ref{fig:grb070721B_dla_1}. 
\tablefoottext{c} The column densities, derived from the EW in the optically thin limit, are considered
as lower limits because the lines are saturated.
}
\label{tab:dla_linelist}
\end{table*}

Because of the small impact parameter of the DLA counterpart ($1''$), the GRB line-of-sight can be
extinguished by the dust in the DLA galaxy if it contains dust (see \citealt{Fynbo2011a} for
an example). \citet{Fynbo2009a} derive the spectral slope between
the optical and X-ray bands, $\beta_{\rm ox}$, to be 0.72. 
According to \citet{Jakobsson2004a}, a dust-extinguished optical afterglow
would result in $\beta_{\rm ox}<0.5$, which is not the case for GRB
070721B. \citet{Zafar2011a} obtained a visual extinction of
$A_V=0.20\pm0.02\,\rm mag$, assuming that all extinction is attributed to an absorber in the
host galaxy of GRB\,070721B. This value is typical for GRB host galaxies \citep{Kann2010a,
Schady2010a, Zafar2011a}. Motivated by the fact that some GRBs show negligible host
reddening, the presence of a foreground absorber cannot be excluded. Assuming the DLA to be
responsible for the observed reddening, we expect that $A_V$ increases only slightly \citep[][their
Fig. 3]{Greiner2011a}. Hence, we can in this case exclude the presence of significant 
amounts of dust, i.e. $A_V<0.20$, in the outskirts of the DLA galaxy.

In addition, we measure the $EW$s of absorption lines in the DLA galaxy spectrum D3 and of
absorption lines imprinted from the DLA galaxy on the GRB afterglow spectrum. The $EW$s and
column density estimates are listed in Table \ref{tab:dla_linelist}. By comparing the two
sets of data it is evident that the DLA galaxy shows a higher $N\left(\rm{\ion{H}{i}}\right)$
than in the afterglow spectrum (see Fig. \ref{fig:grb070721B_dla_1}). This trend is also confirmed
by all the metal $EW$s being higher for the DLA galaxy, despite the individual large errors.
These results are easily explained if most of the DLA galaxy light arises from the central, i.e.
denser, region of the galaxy, while the GRB afterglow radiation shone through a more peripheral
region of the host. 

Finally, the most constraining metallicity estimate for the D3 galaxy is $\left[\rm{Fe}/\rm{H}
\right]>-0.21$ (measured from the GRB afterglow spectrum).  Iron being a very refractory element, a significant
amount of Fe could be depleted into dust grains and not observed in the gas-phase. If this is the
case, then the intrinsic metallicity would increase to solar or super-solar values, while GRB DLAs
typically show solar or sub-solar metallicities \citep[see][and references therein]{Fynbo2009a}.
A more likely scenario is a limited amount of dust in the DLA galaxy, also supported by the low
reddening along the line-of-sight ($A_V<0.20$) estimated from the SED. Thus, the metallicity along the
GRB line-of-sight, i.e. 7.9 kpc away from the galaxy centre, is close to $\approx0.6\,Z_\odot$.
A higher metallicity could possibly be found in the central regions of the DLA galaxy, given the likely
metallicity gradient, but cannot be constrained from our dataset due to absorption-line saturation.
  
In conclusion, the properties of the DLA galaxy D3 are very typical for LBGs,
except that the halo of the galaxy harbours a large amount of neutral hydrogen. This is consistent with
the model that DLAs are gaseous halos of faint high-$z$ LBGs \citep{Fynbo1999a,
Moller2002a,Fynbo2008a,Rafelski2011a}.

\subsubsection{\ion{Mg}{ii} absorbers towards GRB\,050820A}

In addition to the intervening sub-DLA towards GRB\,050820A, the afterglow traverses
two strong \ion{Mg}{ii} absorbers at $z=0.6915$ and 1.4288 (see Table \ref{tab:metal_sample}).
We find four galaxies within 100 kpc that are consistent with either $z=0.6915$,
namely B2, B3 and B5 (Figs. \ref{fig:GRB050820A_B02}, \ref{fig:GRB050820A_B03},
\ref{fig:GRB050820A_B05}, Table~\ref{tab:candidates}), or $z=1.4288$,
namely B6 (Fig. \ref{fig:GRB050820A_B06}, Table~\ref{tab:candidates}).\footnote{The photometric
redshift of B6 also matches that of a weak \ion{Mg}{ii} absorber at $z=1.6204$ towards GRB 050820A.
It is unlikely that the galaxy is the counterpart of the weak \ion{Mg}{ii} absorber,
because the correlation length of weak \ion{Mg}{ii} absorbers is $\sim2$ kpc \citep{Ellison2004a}.}
The properties
of the best-fit galaxy templates are summarised in Table~\ref{tab:grb050820A_mgii}.\footnote{
We remind the reader that the solution of B5 is not unique, because it is only detected in five filters.}

The impact parameters vary between 26 and 88 kpc for the candidates of the \ion{Mg}{ii}
absorber at $z=0.6915$. Galaxy counterparts with an absorption cross-section of more
than 88 kpc are rare; for instance \citet{Chen2010b} used a set of 94 galaxies
($\left\langle z\right\rangle=0.24$) that are located up to 120 kpc around 70
background quasars ($z_{\rm QSO}>0.6$) to study which kind of \ion{Mg}{ii} absorber
could be found in the quasar spectrum arising from these galaxies.
They did not find any strong \ion{Mg}{ii} absorber in a quasar
spectrum that is related to a galaxy at a distance of more than $\sim40\,\rm kpc$ from the line of sight of the quasar.
Hence, the early-type galaxy B2 is probably not the
galaxy counterpart of the absorption-line system. The impact parameters of
B3 and B5 are typical for strong \ion{Mg}{ii} absorbers \citep{Chen2010b}, though at the upper end
of the impact parameter distribution. The
properties of B3 and B5 are very different from each other (Table~\ref{tab:grb050820A_mgii}),
B3 is a massive galaxy with
a high $SFR$, while B5 is a young low-mass galaxy with low $SFR$. 
These differences are also reflected in their luminosities; adopting
the \citet{Dahlen2005a} LF, B3 is $0.2\,L_*$, and B5 is 6 times fainter
($0.03\,L_*$). Both values are in the range of the observed luminosities 
of the \citet{Chen2010b} sample, but in their fainter half.

Galaxy B6 is the most likely galaxy counterpart of the strong \ion{Mg}{ii}
absorber at $z=1.4288$.  The impact parameter of 42 kpc (4\farcs9;
Table~\ref{tab:grb050820A_mgii}) is similar to the impact parameter of the galaxy
counterpart candidates of the \ion{Mg}{ii} absorber at $z=0.6915$.
The properties
of the best-fit galaxy template, displayed in Table \ref{tab:grb050820A_mgii},
show that it is a fairly bright ($0.9\, L_*$) galaxy  with 
$SFR=1.5\,M_{\sun}\,\rm{yr}^{-1}$ (Fig.~\ref{fig:GRB050820A_B06}).

Without additional spectroscopic data it is not possible to decide which of two candidates
is the galaxy counterpart of the strong \ion{Mg}{ii} absorber $z=0.6915$ and
if B6 is the galaxy counterpart to the \ion{Mg}{ii} absorber at $z=1.4288$. \citet{Chen2011a}
places an upper limit on the brightness of the galaxy counterpart of $F775W=27.5\,\rm mag$,
i.e. fainter than a $0.03\,L_*$ galaxy, if the impact parameter is less than 3\farcs5.

\subsection{Limits on galaxy counterparts}

In the previous sections we discussed the properties of the detected
DLA galaxy towards GRB 070721B and of the galaxy counterpart candidates of the
two strong \ion{Mg}{ii} absorbers towards GRB 050820A. In this section
we present the limits, such as luminosity and $SFR$, on the non-detected
galaxy counterparts. The limits on the luminosities shown in Table
\ref{tab:nondetections} are related to the knee of the
\citet{Dahlen2005a} and the \citet{Reddy2008a} LFs. The galaxy counterparts
are fainter than a $0.5\,L_*$ and $1.5\,L_*$ galaxy in all fields, if the impact
parameter is 50 and 100 kpc, respectively. The most stringent upper limit of
$0.1\,L_*$ can be placed on the galaxy counterparts of the intervening sub-DLA
towards GRB 050820A. For this field, \citet{Chen2011a} reports that the galaxy
counterpart is fainter than $F775W=27.5\,\rm mag$, corresponding to $0.06\,L_*$,
if the impact parameter is smaller than 3\farcs5. 

\begin{table}
\caption{Limiting magnitudes of the galaxy counterparts of the intervening absorption line systems}
\centering
\tiny
% \begin{tabular}{l@{\hspace{0.1cm}}c@{\hspace{0.1cm}}c@{\hspace{0.1cm}}c@{\hspace{0.1cm}}c@{\hspace{0.1cm}}c@{\hspace{0.1cm}}}
\begin{tabular}{lcccc}
\toprule
\multirow{2}*{Absorber}	& \multirow{2}*{$z_{\rm abs}$}	& $F775W_{\rm UL}\,\left(\rm{mag}\right)$ for	& \multirow{2}*{$L/L_*$}	& Fynbo \\
			&				& $\theta<50$ (100) kpc				&				& model \\
\midrule
\multicolumn{4}{l}{\textbf{GRB\,050730 $\left(z_{\rm GRB} = 3.969\right)$}}\\
\midrule
DLA			& 3.56439			& $>25.7/>24.6$					& $<0.5/<1.5$ 			& $<0.04$	\\
sub-DLA			& 3.02209			& $>25.7/>24.6$					& $<0.4/<1.1$			& $<0.04$	\\
\ion{Mg}{ii}	& 2.25313				& $>26.5/>25.1$ 				& $<0.2/<0.5$			&	 	\\
\ion{Mg}{ii}	& 1.7743				& $>26.5/>25.1$ 				& $<0.1/<0.5$			&	 	\\
\midrule
\multicolumn{4}{l}{\textbf{GRB\,050820A $\left(z_{\rm GRB} = 2.615\right)$}}\\
\midrule
sub-DLA			& 2.3598			& $>26.7/>26.7$\tablefootmark{a}		& $<0.1/<0.1$ 			& $<0.1$	 	\\
% weak \ion{Mg}{ii}	& 1.6204			& $>26.2/26.2$					& $<0.1/<0.1$			&	 	\\
\midrule
\multicolumn{4}{l}{\textbf{GRB\,050908 $\left(z_{\rm GRB} = 3.3467\right)$}}\\
\midrule
sub-DLA			& 2.6208			& 						& $<0.4/<0.6$ 			& $\gtrsim0.5$	\\
\ion{Mg}{ii}		& 1.5481			& \multirow{-2}*{$>25.0/>24.7$\tablefootmark{b}}& $<0.3/<0.4$			&		\\

\bottomrule
\end{tabular}
\tablefoot{The limiting magnitudes are corrected for Galactic extinction.
		  We adopted the \citet{Dahlen2005a} and \citet{Reddy2008a} LFs to
		  relate an absolute magnitude with the knee of a corresponding
		  LF, $L_*$:
		  $1.59< z < 1.9: M_*=-20.69\pm0.26\,\rm{mag}\,\left(\rm{U-band}\right)$,
		  $1.9\leq z<2.7: M_*=-21.01\pm0.38\,\rm{mag}\,\left(1700\,\rm{\AA}\right)$,
		  $2.7\leq z<3.4: M_*=-20.84\pm0.12\,\rm{mag}\,\left(1700\,\rm{\AA}\right)$.
		  We extended the range of the LFs from $z=1.59$ to 1.50 and from $z=3.4$ to 3.56,
		  because the change in the $M_*$ is smaller than its uncertainty. To compute
		  the absolute magnitude we used: $M=m-DM\left(z\right)+2.5\log\left(1+z\right)$,
		  where $DM\left(z\right)$ is the distance modulus. Different filters than $F775W$
		  were used if a filter was closer to the central wavelength of the LF than the
		  $F775W$ band. In the column "Fynbo model"  we display the limits on the
		  luminosity of DLA galaxies based on the model by  \citet{Fynbo2008a}.
	  \tablefoottext{a} The $F625W$ magnitude was used.
	  \tablefoottext{b} The $R$-band magnitude was used.
	  }
\label{tab:nondetections}
\end{table}

\citet{Fynbo1999a} suggested that DLAs are the gaseous halos of LBG
\citep[see also][]{Moller2002a, Fynbo2008a, Rafelski2011a}. To confront our
measurements with this model, we use the predictions on the brightness
of galaxy counterparts by \citet{Fynbo2008a}. In their model, simple
and constrained scaling relations of galaxies in the local Universe are
used to predict the luminosity of intervening DLAs based on their
metallicity and their impact parameter. These relations are strictly speaking
only valid for $z\sim3$ DLAs. We also adopt this model for DLAs at lower
redshift and for sub-DLAs. The main difference between a sub-DLA and a DLA is
a larger covering fraction, i.e. impact parameter, if DLAs and sub-DLA are drawn
from the same population of galaxies. Based on the measured metallicities, we estimate
the brightness and the impact parameter of the unidentified DLA galaxies
towards GRBs 050730 and 050820A. For GRB 050908, we used
$EW_{\rm rest}\left(\rm{\ion{Si}{ii}}\,\lambda1526\right)$ as a metallicity proxy
\citep{Prochaska2008a}.

The estimated brightnesses, shown in Table \ref{tab:nondetections}, indicate
that most galaxy counterparts evaded detection because of their intrinsic faintness.
This result is not surprising. The majority of all $z\sim3$ DLA galaxies are
expected to be fainter than $\sim29.5$ mag in the $R$-band, i.e. $L\lesssim0.012\,L_*$,
in the \citet{Fynbo2008a} model. The model clearly rules out large impact parameters.
Even the larger covering fraction of sub-DLAs does not imply impact parameters
of more than 2--3$''$ (15.7--23.5 kpc at $z\sim3$). The intervening DLA towards
GRB 050908 possibly has a $\sim0.5\,L_*$ galaxy counterpart.

We find several objects in the field of GRBs 050730 and 050908 that fulfil both criteria,
being very faint and having a very small impact parameter (Table \ref{tab:candidates}).
There are several objects with a brightness of $F775W\sim28.4\,\rm mag$ within 2\farcs5
of the afterglow position of GRB 050730. If one of these objects is the galaxy
counterpart to the sub-DLA, it is a $0.03\,L_*$ galaxy. In the field of GRB
050908 we find four objects ranging in brightness from $R=25.3$ to 27\,mag within
$4''$ of the afterglow position, corresponding to $L\lesssim0.4\,L_*$. To elucidate the
nature of these objects, more data is required. 

\subsection{Quasar radiation field}\label{sec:disc_qso}

\begin{table}
\caption{Absorption and emission lines of QSO J1408-0346 and the strong \ion{Mg}{ii} absorber at $z=1.77425$ towards it.}
\centering
\begin{tabular}{rrccc}
\toprule
\multicolumn{1}{c}{$\lambda_{\rm obs}$}	&\multicolumn{1}{c}{$\lambda_{\rm rest}$}	& Species	& Redshift	& $EW_{\rm rest}$\tablefootmark{a}\\
\multicolumn{1}{c}{(\AA)}		&\multicolumn{1}{c}{(\AA)}			&		& 		& (\AA)\\
\midrule
3658				& 911.20			& Ly-limit				& 3.01		\\
$\mathbf{\approx 4168.8}$	& $\mathbf{\approx 1033}$	& \textbf{Ly$\beta$ + \ion{O}{vi}}	& $\mathbf{\approx 3.04}$\\
\textbf{4888}			& \textbf{1215.80}		& \textbf{Ly$\alpha$}			& \textbf{3.02}		\\
\textbf{4992.72}		& \textbf{1239.42}		& \textbf{\ion{N}{v}}			& \textbf{3.028}		\\
%5060.41				& 1260.40			& \ion{Si}{ii}				& 3.015		\\
%5263.65				& 1309.30			& \ion{Si}{ii}$^{*}$			& 3.020		\\
5621.30				& 2026.14			& \ion{Zn}{ii}	 			& 1.77425	&$< 0.73$\\
6208.75				& 1548.20			& \ion{C}{iv} 	 			& 3.010	&\\
6210.62				& 1548.20			& \ion{C}{iv}  				& 3.011	&\\
6219.37				& 1550.77			& \ion{C}{iv}  				& 3.010	&\\
6220.87				& 1550.77			& \ion{C}{iv}  				& 3.011	&\\
\textbf{6229.35}		& \textbf{1545.86}		& \textbf{\ion{C}{iv}}			& \textbf{3.030}		&\\
6503.52				& 2344.21			& \ion{Fe}{ii} 				& 1.77425	&$<0.49$\\
6610.51				& 2382.77			& \ion{Fe}{ii} 				& 1.77425	&$0.84\pm0.15$\\
7145.59				& 2586.65			& \ion{Fe}{ii} 				& 1.77425	&$<0.54$\\
7213.61				& 2600.17			& \ion{Fe}{ii} 				& 1.77425	&$0.62\pm0.19$\\
$\mathbf{\approx 7674.8}$	& $\mathbf{\approx 1910}$	& \textbf{\ion{Si}{ii}, \ion{C}{iii}/\ion{Fe}{ii}}& \textbf{3.018}\\ 
7756.18				& 2796.35			& \ion{Mg}{ii}				& 1.77425	&$1.24\pm0.34$\tablefootmark{b}\\
7777.52				& 2803.53			& \ion{Mg}{ii}				& 1.77425	&$1.25\pm0.19$\\

\bottomrule
\end{tabular}
\tablefoot{The table summarises the most important absorption and emission lines of the quasar
    and the strong \ion{Mg}{ii} absorber at $z=1.77425$. We show only those absorption lines that
    are redward of the quasar Ly$\alpha$ emission line. Emission lines are displayed in bold.\\
    \tablefoottext{a} The restframe $EW$s ($1\sigma$ errors) and the $3\sigma$ detection limits are
    measured assuming an aperture of $333\,\rm{km\,s}^{-1}$, i.e. $2\times FWHM$
    a Gaussian with Doppler parameter $b=100\,\rm{km\,s}^{-1}$. \tablefoottext{b} The $EW$ of the lines
    of the \ion{Mg}{ii} doublet are similar, indicating that the lines are saturated.}
\label{tab:qso_linelist}
\end{table}

\begin{figure*}
\centering
\includegraphics[bb= 2 2 1033 632,clip, width=.92\textheight, angle=90]{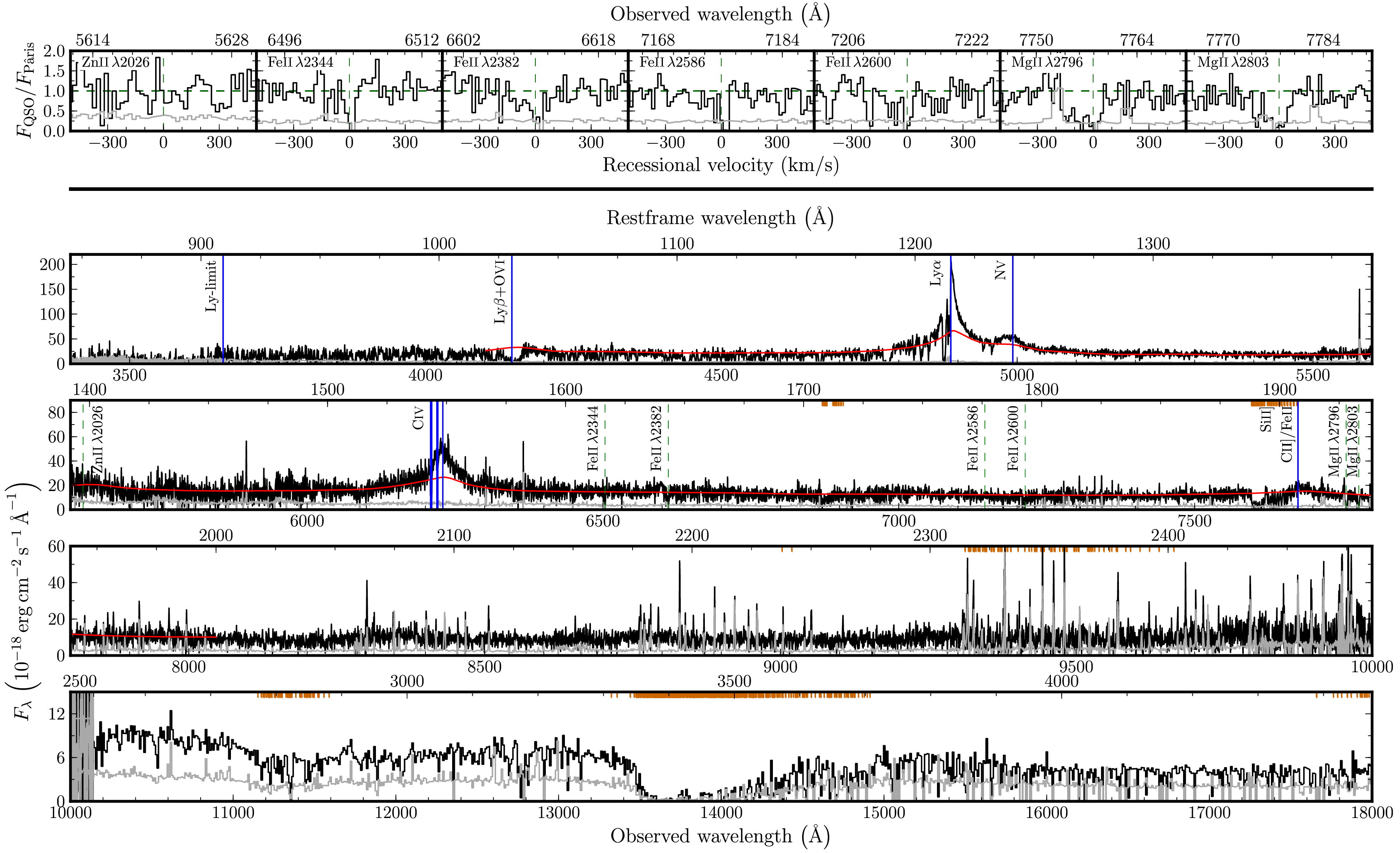}
\caption{X-shooter spectrum of QSO J1408-0346 (bottom). The spectrum beyond $1.8\,\mu\rm m$ is not shown due
	to uncertainty in the flux calibration. The observed and rest-frame wavelengths are shown below
	and above each panel, respectively. The error spectrum is shown in grey. Regions
	of strong telluric features (atmosphere transparency $< 20\%$) were not used in the spectral
	analysis and are marked by small boxes at the top of each panel (NIR: Gemini observatory).
	The red line is the average $z = 3$ QSOs spectrum taken from \citet{Paros2011a}. Prominent
	absorption and emission lines from the quasar are indicated by vertical solid lines,
	and the absorption lines from the strong \ion{Mg}{ii} absorber at $z = 1.77425$ by vertical
	dashed lines (see Table~\ref{tab:qso_linelist}). Zoom-ins on the absorption lines of the \ion{Mg}{ii}
	absorber are shown in the top panel. They were normalised with the QSO composite spectrum.
	}
 \label{fig:qso_spectrum}
\end{figure*}

In Sect.~\ref{sec:grb050730}, we alluded to the serendipitous coincidence of the sub-DLA
($z=3.02$) towards GRB\,050730 and the QSO that is 17\farcs 5 south of the
GRB position. Before we discuss their connection, we present general properties of the quasar.

After subtracting the bright $R=18.2\,\rm mag$ foreground star (Sect.~\ref{subsec:reduction};
Table~\ref{tab:candidates}), we measure an $R$-band magnitude of $20.78\pm0.05$ mag
(corrected for Galactic extinction, Table~\ref{tab:candidates}). The coordinates of the quasar are
$RA\rm{(J2000)}=14^{\rm h}08^{\rm m}17\fs135$ and $DEC\rm{(J2000)}=-03^\circ46'35\farcs35$,
with an uncertainty of 0\farcs4. Hereafter the QSO is denoted as QSO J1408-0346.
Figure \ref{fig:qso_spectrum} shows the quasar spectrum secured 
with X-shooter on 18 April, 2010. Table~\ref{tab:qso_linelist} summarises the detected quasar
emission and absorption lines; the mean emission line redshift is $z_{\rm em} = 3.023\pm0.007$.  
The emission lines are substantially narrower in comparison to the quasar composite
spectrum by \citet{Paros2011a}, which is overplotted in red. 
In addition to the emission
lines, we detect \ion{C}{iv} and
Ly$\alpha$ in absorption at $z_{\rm qso}$ (Table~\ref{tab:qso_linelist}). 
The narrow \ion{C}{iv} absorption lines consist of two blueshifted velocity components
at 1420 and 1497 km s$^{-1}$ with respect to the maximum of the
\ion{C}{iv} emission line.

The proximity of the GRB line-of-sight can constrain the quasar radiation
field. \citet{Fox2007a} reports that DLAs and sub-DLAs bearing \ion{N}{v} are
extremely rare, and that several of these have a nearby QSO coincident
($>33\%$) at the (sub)-DLA redshift \citep[see also][]{ Fox2007a,
Ellison2010a}. We do not find evidence for \ion{N}{v} in absorption in the GRB
afterglow spectrum. Besides this, the ionising UV radiation field of a bright
quasar can also lead to visible underdensity in the Ly$\alpha$ forest as the
redshift approaches the quasar emission redshift. This is referred to as the
proximity effect \citep{Carswell1982a, Tytler1987a, Bajtlik1988a}. In addition,
if the quasar radiates isotropically, the line of sight of GRB\,050730 might show
an underdensity in the Ly$\alpha$ forest around the redshift of the
foreground quasar \citep[i.e. the transverse proximity
effect;][]{Bajtlik1988a}.  At the projected distance of $137$ kpc, the QSO with
an inferred Lyman-limit flux density of $9\,\mu\rm{Jy}$ exceeds the \ion{H}{i}
UV background at $z\sim3$ by a factor of $\sim170$ \citep{Guimaraes2007a,
DallAglio2009a}. In this case, the Ly$\alpha$ forest at the redshift of the
QSO is dominated by the sub-DLA so it impossible to infer if there is an effect
of the QSO on the IGM along the GRB afterglow line-of-sight.

\subsection{Correlated structures}\label{sec:qso_subdla}

The spectra of GRB\,050730 and the QSO 17\farcs5 south of it
(Fig. \ref{fig:fov_050730}) share a common wavelength interval,
as do GRB\,070721B and the bright galaxy D7 20\farcs7 SW
of it (Fig. \ref{fig:fov_070721B}). These allow us to identify
correlated structures in both fields. We are in particularly
interested in absorbers with metal lines, because they are
related to galaxies. Their detections provide information on
galaxy impact parameters and their clustering.

Correlated structures have been found in several quasar pairs on a
length scale of a few kpc up to $\sim100$ kpc \citep[e.g.][]{Smette1995a,
Petry1998a, Ellison2004a, Ellison2007a}, and on a length scale of
several hundreds of kpc up to a few Mpc \citep{Francis1993a, Francis1996a,
Fynbo2003a, DOdorico2002a, Coppolani2006a}. In the literature, two different
scenarios are discussed to explain correlated structures on these different
length scales: a) the lines of sight probe the halo of one galaxy, or of a
galaxy and a satellite galaxy, and b) the different lines of sight probe
different galaxies that belong to a group of galaxy and are part of a large
scale structure. Here we argue that the latter interpretation of correlations
being due to large scale structure explains better our observations of the
absorbers in both GRB fields.

\subsubsection{The field of GRB\,050730}

The lines of sight of GRB 050730 and QSO J1408-0346 are separated by 17\farcs5.
We identify a \ion{Mg}{ii} at $z\approx1.774$ in both of them, as well as
the sub-DLA towards the GRB and QSO J1408-0346 that are at almost identical
redshifts ($z_{\rm sub-DLA}= 3.02209$, $z_{\rm QSO,\, em} = 3.023$).

The coincidence of the sub-DLA and the QSO is in line with evidence for anisotropic
emission of QSOs, which seem to have cleared their surroundings preferentially
towards us rather than perpendicular to the line-of-sight \citep{Hennawi2007a}.
At least it suggests that QSOs reside in over-dense regions \citep{Rollinde2005a,
FaucherGiguere2008a}. The angular distance of 17\farcs5 between the sub-DLA and
the QSO translates into a transverse distance of 137 kpc at $z\simeq3.022$, which
is several times larger than the typical sub-DLA impact parameter of $\sim39$ kpc
in \citet{Peroux2011a} and \citet{Krogager2012a}. \citet{Ellison2010a} argue that proximate DLAs,
i.e. DLAs within a comoving distance of 42 Mpc ($\Delta v=3000\,\rm{km}\,\rm{s}^{-1}$)
around QSOs, are not associated with the QSO host, but rather sample overdensities
around it \citep[see also][]{Russell2006a}.  Therefore, the intervening sub-DLA towards
GRB\,050730 could be an overdensity in the vicinity of the QSO rather than a part of the
massive halo of the quasar host galaxy.

The common \ion{Mg}{ii} absorbers towards GRB\,050730 and the QSO are at $z_{\rm
GRB, \rm{\ion{Mg}{ii}}}=1.7731$ and $z_{\rm QSO, \rm{\ion{Mg}{ii}}}=1.77425$.
\citet{Vergani2009a} reported a restframe
\ion{Mg}{ii} equivalent width of $0.93\pm0.03\,\rm\AA$ and \citet{Prochaska2007a} reported
that the doublet is saturated so that the absorber can be placed in the strong
\ion{Mg}{ii} absorber category. For the intervening
\ion{Mg}{ii} absorber towards the QSO, we measure $EW_{\rm
rest}(2796\,\rm{\AA})=1.24\pm0.34$ (saturated;
Table~\ref{tab:qso_linelist}). In addition, we detect \ion{Zn}{ii} $\lambda$
2026 and several \ion{Fe}{ii} absorption lines of that absorber at the same redshift,
summarised in Table~\ref{tab:qso_linelist}.
At the redshift of $1.774$, the angular distance of 17\farcs5 translates into a
projected distance of $150$ kpc, while their redshift implies a velocity distance
of 124 km s$^{-1}$. \citet{Chen2010b} reports a typical impact parameter of around $\sim30$ kpc 
for \ion{Mg}{ii} absorbers \citep[see also][]{Smette1995a}, arguing against the
hypothesis that both strong \ion{Mg}{ii} absorbers are associated with the same
galaxy.\footnote{The galaxy counterpart of the \ion{Mg}{ii} absorber towards the QSO 
is fainter than 26.5 (25.1) mag in $F775W$ (corrected for Galactic extinction)
for an impact parameter of 50 kpc (100 kpc), assuming that the galaxy counterpart is
neither in the glare of the QSO, nor at the position of the foreground star.
Adopting the \citet{Dahlen2005a} LF, the limiting magnitudes correspond to 
$L<0.04\,L_*$ and $L<0.16\,L_*$.
 }
Finally, we note that there is a high-column density absorber at $z\approx2.98$ in
the afterglow spectrum, which has no counterpart in the spectrum of the quasar.

\subsubsection{The field of GRB\,070721B}

In Sect. \ref{sec:res_070712B} we identified object D3 as the galaxy counterpart
of the intervening DLA in the afterglow spectrum of GRB 070721B. In addition, we
identified another intervening DLA towards the Ly$\alpha$ emitter D7 that is at
a similar redshift and has a similar \ion{H}{i} column density to the DLA galaxy,
but is 160.7 kpc from the location of D3 (Figs. \ref{fig:grb070721B_dla_1}, \ref{fig:fov_070721B}).
It is unlikely that the intervening DLA towards D7 is gravitationally bound to
the DLA galaxy D3 because of this large impact parameter. The average impact
parameter of DLAs is 13 kpc, although \citet{Francis1996a} also report a candidate 
a counterpart with impact parameter of 182 kpc.

D3 and the intervening absorber
towards D7 appear more similar to the pairs of intervening absorption-line systems
discussed in \citet{Ellison2007a}.\footnote{The galaxy counterpart
of the intervening DLA towards D7 is fainter than 25.2 mag in $F775W$, assuming an impact
parameter of 50 to 100 kpc, corresponding to $L\lesssim0.7\,L_*$, using the LF in
\citet{Reddy2008a}.}
 They found found a pair of sub-DLAs at $z=2.94$ and a pair
of DLAs at $z=2.66$ towards a binary quasar, where the individual absorbers
of each pair are separated by $\sim100\,\rm kpc$. Based on simulations,
they showed that the presence of a large scale structure is more likely
than the different lines of sight probing the halo of two massive galaxies
at $z=2.66$ and 2.94, respectively. In this case, the probability increases from
$p\lesssim 10^{-3}$ to 0.01.
We find several correlated structures on a length scale of $\sim160\,\rm kpc$
in the fields of GRBs 050730 and 070721B. These are most likely further
examples for a group environment of intervening absorption-line systems.

\section{Conclusions}\label{sec:conclusion}

The aim of our work is to detect galaxy counterparts of high-$z$ intervening sub-DLAs and
DLAs. In contrast to previous studies, we use GRBs to have a clear view on the region that
is usually outshone by the glare of a quasar. Since the launch of the \swift \ satellite, seven
intervening sub-DLAs and DLAs have been found towards six GRBs. Among them four lines-of-sight have
sufficient photometric and spectroscopic data to study very faint objects.

In our study we successfully detected the DLA galaxy, causing the intervening DLA absorption
towards GRB\,0707021B, $1''$ from the afterglow position, as suggested by \citet{Chen2009a}
and \citet{Fynbo2009a}. However, these authors did not present any spectroscopic evidence for
their inference. In fact, the DLA galaxy would have been almost impossible to detect by direct
imaging if the background source would have been a 19-mag quasar. This underlines the argument
by \citet{Jakobsson2004a} that studies of the galaxy counterparts of intervening absorption
line systems towards QSOs can be affected by misidentifications. Hence, proximity is not
sufficient for an association. 
For instance in the case of GRB 050820A, the extensive photometric and spectroscopic campaigns
allow us to successfully rule out all objects brighter 26.2 mag in $F625W$-band  within 3\farcs7,
corresponding to $L\lesssim0.1\,L_*$ at $z=2.3598$. 
Assuming that sub-DLAs and DLAs are LBGs weighted by their \ion{H}{i} cross-section, \citet{Fynbo2008a}
showed that
the overwhelming majority of intervening DLA galaxies are expected to be fainter than this, naturally
explaining the non-detections. Even
with the largest telescopes, it is difficult to detect and elucidate the nature of such faint
objects. On the other hand, there are cases where intervening absorption-line systems were successfully
associated with galaxies at larger distances \citep[e.g.][]{Francis1993a},
however they are just a minority.

The disadvantage of our approach is the vast amount of observing time required. Deep
multi-filter observations exist for several GRBs. However, similar extensive
spectroscopic campaigns do not exist for other GRB fields. In those cases, one relies on
SED fitting techniques. \citet{Rao2011a} showed that this can indeed be successful,
in particular for very faint objects where only emission-line spectroscopy is feasible.
Recently, \citet{Peroux2011a} performed a survey for galaxy counterparts of intervening
DLAs using the integral field unit VLT/SINFONI \citep[see also][]{Christensen2004b, Bouche2011a}.
This approach could be complementary to the classical strategy of performing very deep
imaging campaigns and spectroscopic follow-ups on candidates, or the use of narrow band
and broad band strategy to search for emission line objects \citep[e.g.][]{Vreeswijk2003a}.

The DLA galaxy ($z=3.096\pm0.003$) in the field of GRB\,070721B is the most luminous high-$z$
DLA galaxy known \citep{Chen2009a}, and is almost as distant as the highest-redshift DLA galaxy known
so far at $z=3.15$. The number of high-$z$ sub-DLA and DLA galaxies increases thus from nine to 10.
DLA J0212-0211 
is very metal rich,
$\left[\rm{Fe}/\rm{H}\right] > -0.21$. It does not differ from normal LBGs,
with the exception of the large amount of \ion{H}{i} in its halo.This supports a model in which
DLAs can be gaseous halos of LBGs \citep{Fynbo1999a,Moller2002a,Fynbo2008a,Rafelski2011a}.

The extensive photometric and spectroscopic campaigns allowed us to identify
galaxy counterpart candidates of two strong \ion{Mg}{ii} absorbers at $z=0.6915$ and
1.4288 towards GRB\,050820A. The most likely candidates have impact parameters between
36.8 and 42.0 kpc, this being in the expected range of \ion{Mg}{ii} absorbers \citep{Chen2010b}.
Their properties point to young star-forming galaxies, ranging in luminosity from $0.2\,L_*$ to
$0.9\,L_*$.

Finally, we studied the presence of correlated structures as a particular class of intervening
absorption-line systems, a phenomenon observed in the field of GRBs 050730 and 070721B. We find
evidence for three correlated structures in the field of GRB 050730 and one in the field of GRB
070721B. These absorbers range in redshift from $z=1.774$ to 3.096 and are probed over transverse
distances between 137 and 161 kpc. All of them have associated metal absorption-lines. It is unlikely
that the objects in these correlated structures are gravitationally bound to each other, based on simulations
\citep[e.g.][]{Ellison2007a}. The detection is nevertheless intriguing. The average distance between
quasar pairs is several times larger. Up to now, intervening absorption-line systems with coincidences
in redshift have been found with separations of several hundreds of kpc up to a few Mpc
\citep[e.g.][]{Francis1993a, DOdorico2002a}, but only few with a separation of $\sim150$ kpc \citep{Ellison2007a}.
These serendipitous discoveries are important for studying the absorption cross-section of
intervening absorption-line systems, their correlation length and their implications on galaxy
groups in the early Universe.

\section*{Acknowledgments}
SS, ADC and PJ acknowledge support by a Grant of Excellence from the Icelandic Research Fund.
JPUF acknowledges support form the ERC-StG grant EGGS-278202. AR acknowledges support by the
Th\"uringer Landessternwarte Tautenburg and by the Graduierten-Akademie Jena, Germany.
SS thanks Daniele Malesani (Dark Cosmology Centre, Copenhagen) for many productive and valuable
discussions and David Alexander Kann (Th\"uringer Landessternwarte Tautenburg, Germany) for many
valuable comments.

\appendix
\section{Additional figures}
\newpage
\begin{figure}
\centering
\includegraphics[bb= 11 6 1097 589, clip, width=0.49\textwidth, angle=0]{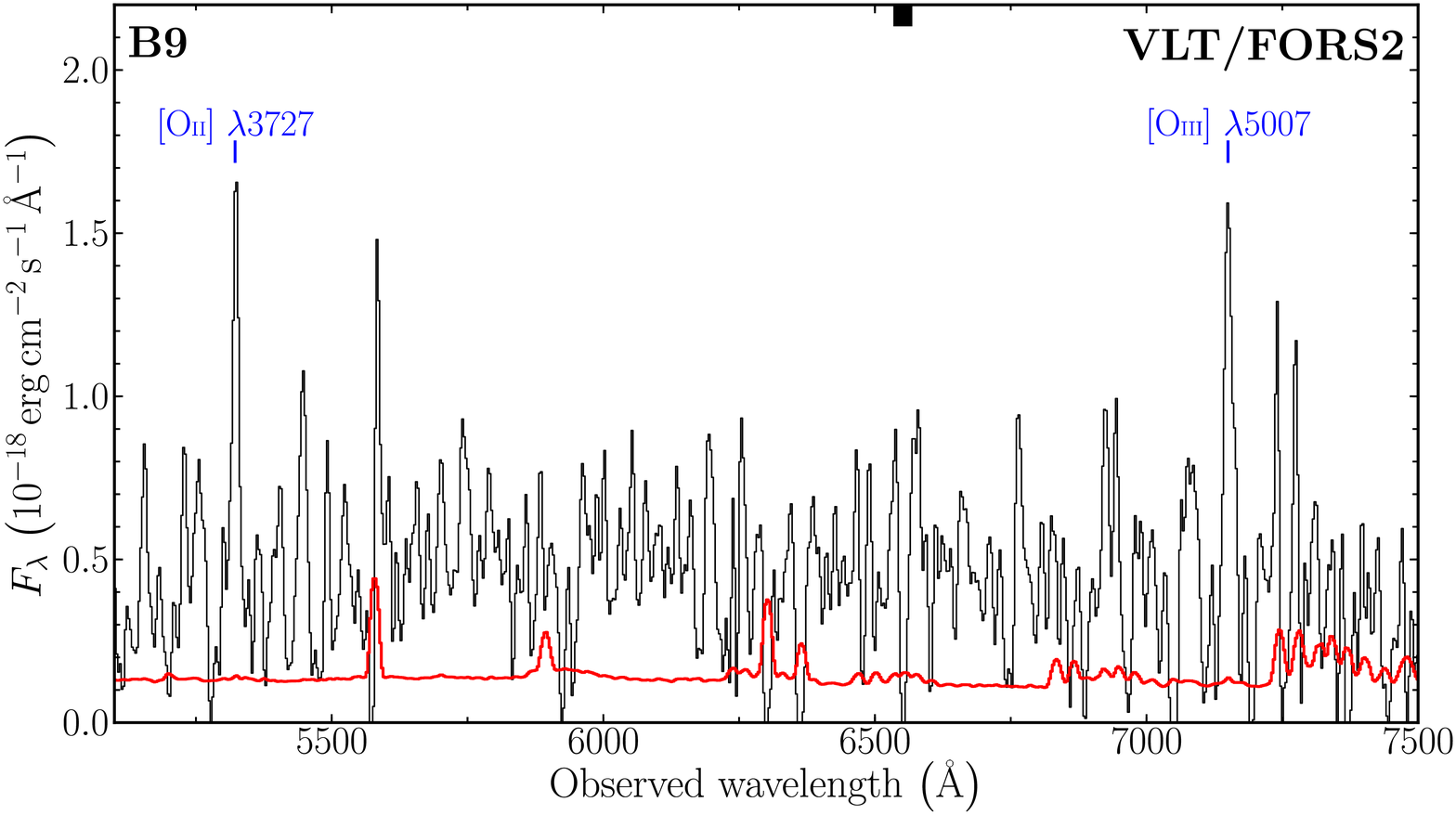}
\caption{
	Smoothed spectrum of the galaxy B9 in the field of view of GRB\,050820A.
	 Based on the identification of [\ion{O}{ii}] $\lambda$3727 and [\ion{O}{iii}] $\lambda$5007
	 the redshift is $z=0.428$. The galaxy is not related to any of the intervening absorbers
	 seen in the afterglow spectrum. The error spectrum is overplotted. Regions of strong telluric
	 features (atmosphere transparency $< 20\%$) were not used in the spectral analysis and are
	 marked by small boxes at the top.
}
 \label{fig:GRB080520A_B09}
\end{figure}

\begin{figure}
\centering
\includegraphics[bb= 10 10 678 551, clip, width=0.49\textwidth, angle=0]{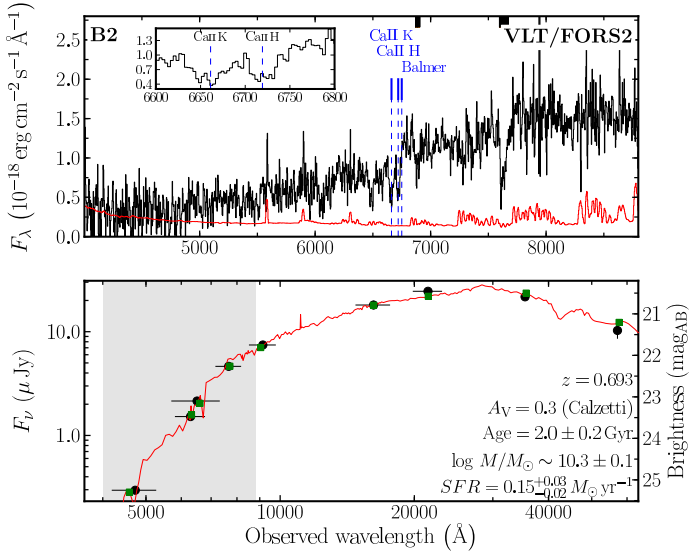}
\caption{Galaxy B2 in the field of GRB\,050820A. \textit{\textbf{Top}}: Spectrum acquired with VLT/FORS2.
	 The identified \ion{Ca}{ii} H\&K absorption lines and the Balmer break are marked. The inset
	 zooms in on the position of the both absorption lines.
	 Based on the identification of the three absorption features the redshift of the galaxy is
	 $z=0.693$.  The error spectrum
	 is overplotted. Regions of strong telluric features (atmosphere transparency $< 20\%$)
	 are not used in the spectral analysis and are marked by small boxes at the top.
 	 \textbf{\textit{Bottom}}: SED from $g'$-band to $5.8\,\mu\rm m$.
	 The observed data points (corrected Galactic extinction) are shown as circles with error bars.
	 The solid line displays the best fit model of the SED ($\chi^2=16.3$; number of filters = 9).
	 The model predicted magnitudes are superposed (squares). The gray area highlights the
	 interval that is covered by the spectrum above.
	}
 \label{fig:GRB050820A_B02}
\end{figure}

\begin{figure}
\centering
\includegraphics[bb= 11 6 1097 589, clip, width=0.49\textwidth, angle=0]{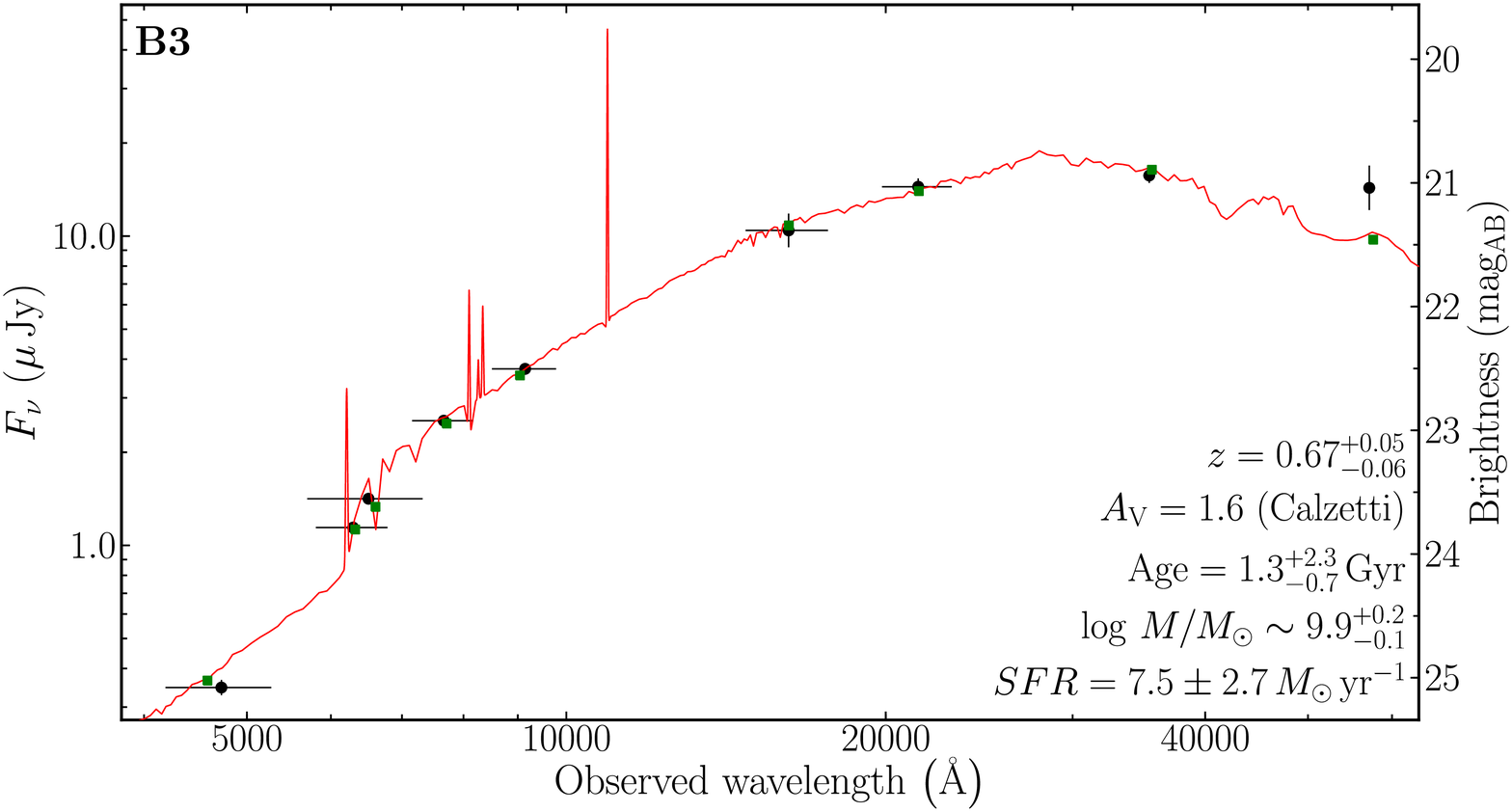}
\caption{Similar to Fig. \ref{fig:GRB050820A_B02}, but for galaxy B3 from $g'$-band to $5.8\,\mu\rm m$. 
		The fit quality is $\chi^2 =7.3$ for 9 filters.
	}
 \label{fig:GRB050820A_B03}
\end{figure}

\begin{figure}
\centering
\includegraphics[bb= 11 6 1097 589, clip, width=0.49\textwidth, angle=0]{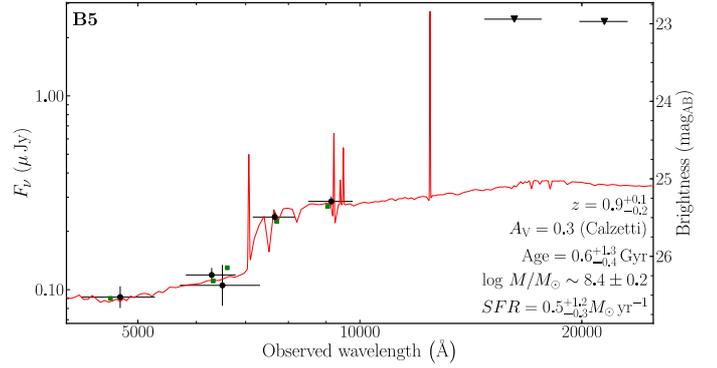}
\caption{Similar to Fig. \ref{fig:GRB050820A_B02}, but for galaxy B5 from $g'$- to $K_s$-band. 
		The fit quality is $\chi^2=1.5$ for 5 filters.
	}
 \label{fig:GRB050820A_B05}
\end{figure}
\newpage
\begin{figure}
\centering
\includegraphics[bb= 11 6 1097 589, clip, width=0.49\textwidth, angle=0]{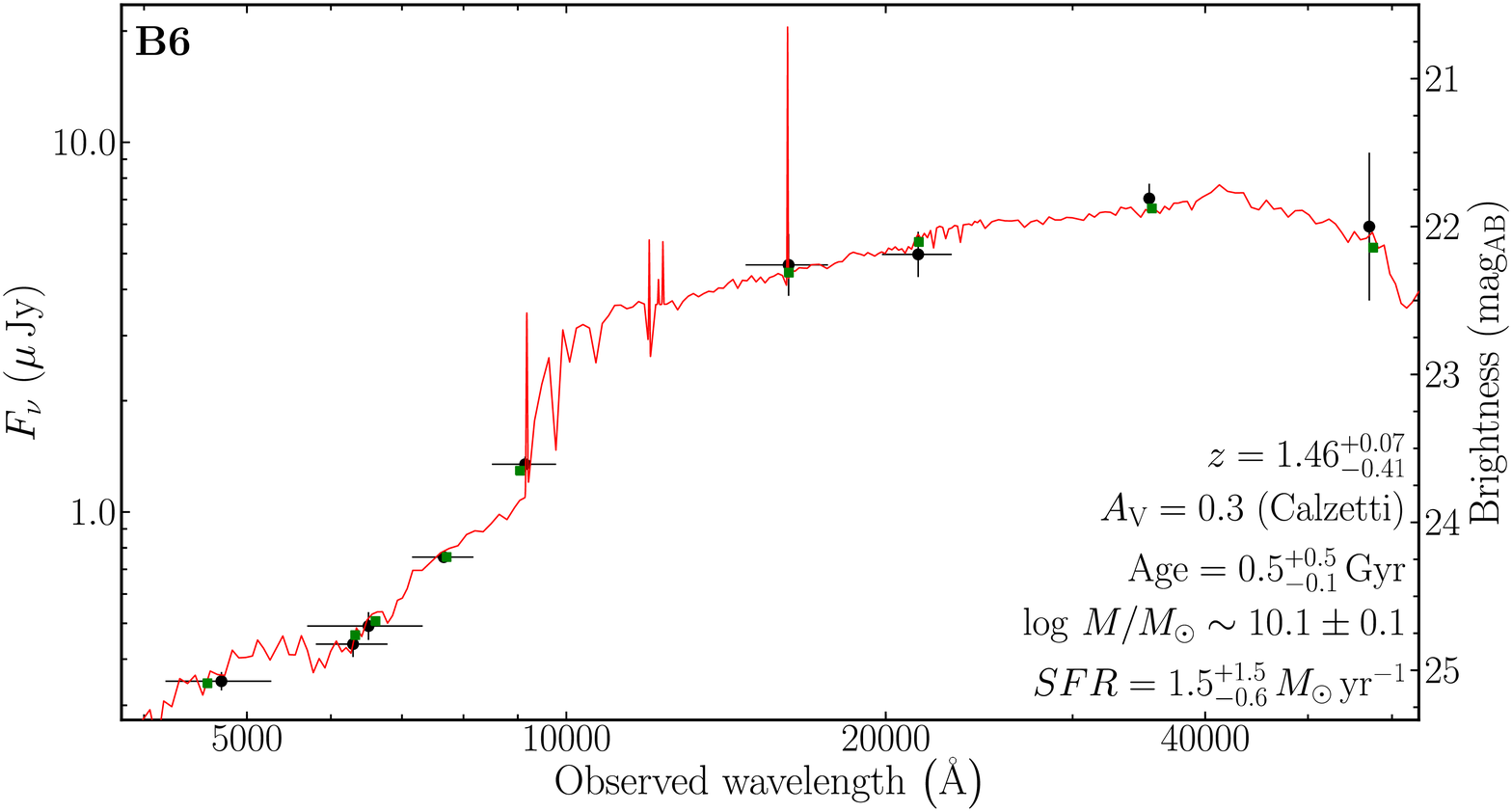}
\caption{Similar to Fig. \ref{fig:GRB050820A_B02}, but for galaxy B6 from $g'$-band to $5.8\,\mu\rm m$.
		The fit quality is $\chi^2 =1.6$ for 9 filters.
 	}
 \label{fig:GRB050820A_B06}
\end{figure}

\label{lastpage}
\end{document}